\documentstyle[aaspp4]{article} 
\begin{document}
\title {HYDRODYNAMICAL EFFECTS IN INTERNAL SHOCK \\OF RELATIVISTIC OUTFLOWS}
\author { MOTOKI KINO \altaffilmark{1}, 
AKIRA MIZUTA\altaffilmark{2}
 and SHOICHI YAMADA \altaffilmark{3}}
\affil {\altaffilmark{1}SISSA, via Beirut 2-4, 34014 Trieste, Italy;}
\affil {\altaffilmark{2}
Institute of Laser Engineering, Osaka University,
Suita, Osaka 565-0871, Japan;}
\affil {\altaffilmark{3}
Science and Engineering, Waseda University, 
Shinjyuku, Tokyo 169-8555, Japan}

\begin{abstract}

We study both analytically and numerically 
 hydrodynamical
effects of two colliding shells, 
the simplified models of the 
internal shock in various relativistic outflows 
such as gamma-ray bursts and blazars.
We pay particular attention to 
three interesting cases:
a pair of shells with the same rest mass density 
(``{\it equal rest mass density}''),
a pair of shells with the same rest mass (``{\it equal mass}''),
and a pair of shells with the 
same bulk kinetic energy (``{\it equal energy}'')
measured in the intersteller medium (ISM) frame.
We find that 
the density profiles are significantly affected by the 
propagation of rarefaction waves.
A split-feature appears 
at the contact discontinuity of two shells
for the ``equal mass'' case, while no significant split appears
for the ``equal energy'' and 
``equal rest mass density'' cases.
The shell spreading with 
a few ten percent of the speed of light 
is also shown as a notable aspect caused by rarefaction waves.
The conversion efficiency 
of the bulk kinetic energy to internal one
is numerically evaluated.
The time evolutions of the efficiency show 
deviations from the widely-used inellastic 
two-point-mass-collision model.

\end{abstract}

\keywords
{galaxies: jet --- gamma rays: bursts --- gamma rays: theory
--- radiation mechanisms: non-thermal --- shock waves}

\section{INTRODUCTION}\label{sec:intro}

The internal shock scenario proposed by Rees (1978) is 
one of the most promising models to 
explain the observational feature of relativistic outflows
as in gamma-ray bursts, and blazars 
(e.g., Rees \& Meszaros 1992; Spada et al. 2001).
In this scenario,
the bulk kinetic energy of the outflowing plasma
is converted into thermal energy and non-thermal 
particle energy by the shock dissipation 
and particle acceleration, respectively,
and explain the large power of these objects.
Based on this scenario, a lot of authors have
attempted to link the observed temporal profiles 
to multiple internal interactions
(e.g., Kobayashi, Piran, \& Sari 1997 (hereafter KPS97); 
Panaitescu, Spada \& Meszaros 1999;
Tanihata et al. 2002; Nakar \& Piran 2002 (hereafter NP02)), 
looking for crucial hints on the central engine of
these relativistic outflows.

Most of the previous works focus on the comparison with the 
observed light curves and model predictions 
employing a simple inelastic 
collision of two point masses (KPS97) 
and little attention has been paid  
to hydrodynamical processes in the shell collision. 
However,
it is obvious that, 
in the case of relativistic shocks,
the time scales in which 
shock and rarefaction waves cross the shells
are comparable to 
the dynamical time scale $\Delta^{'}/c$,
where $\Delta^{'}$ is the shell width measured in 
the comoving frame of the shell
and $c$ is the speed of light.
Since the
time scales of observations of these relativistic outflows 
(e.g., Takahashi et al. 2000 for blazar jet;
Fishman \& Meegan 1995 for GRBs)
are much longer than the dynamical time scales,
the light curves should contain the footprints of 
these hydrodynamical wave propagations.
Thus, it is very interesting
to clarify the  difference between 
the simple two-point-mass-collision 
(hereafter two-mass-collision) model
and the hydrodynamical treatment.
The recent study by
Kobayashi \& Sari 2001 (hereafter KS01)
reports that collided shells are reflected
from each other by the thermal expansion.
Since they perform a hydrodymamical simulation
and show the reflection feature for a single case,
the detail of propagations of rarefaction waves 
for various cases of collisions is not discussed.
The aim of this paper
is to clarify the hydrodynamical effects 
including the propagations of rarefaction wave.
As the simplest case,  
we explore the hydrodynamics 
of two-shell-collisions in the  internal shock model.
Since we are mainly interested in 
the hydrodynamical processes themselves, 
it is beyond the scope of this paper to make a detailed comparison 
of the observed phenomena with the model results.

We consider the time evolution of two colliding shells 
in relativistic hydrodynamics
in \S 2.
In \S 3, we discuss the 
application to GRBs and blazars.
The summary and discussion are given in \S 4.

\section{HYDRODYNAMICS}

Here we consider the
hydrodynamics of the two-shell-interactions.
Our intention is to derive analytically 
various time scales
for shocks and rarefaction waves crossing the shells.
The fundamentals of relativistic shocks are 
given by Landau \& Lifshitz (1959) and Blandford \& McKee (1976).
Our main assumptions are 
(1) we adopt a planar 1D shock analysis and
neglect radiative coolings for simplicity,
(2) neglect the effect of magnetic fields,
and
(3)limit our attention to  shells with relativistic speeds.
We are currently planing 2D studies. 
The role of magnetic fields is still under debate. 
A multi-frequency analysis of TeV blazars shows 
that 
the energy density of magnetic field 
is smaller than that of 
non-thermal electrons (Kino, Takahara, \& Kusunose 2002).
As for (3), it is self-evident
that the relativistic regime is most important,
since emissions from GRB and blazars show a substantial
Doppler boost.

\subsection{Shock Jump Condition}

\begin{figure}
\plotone{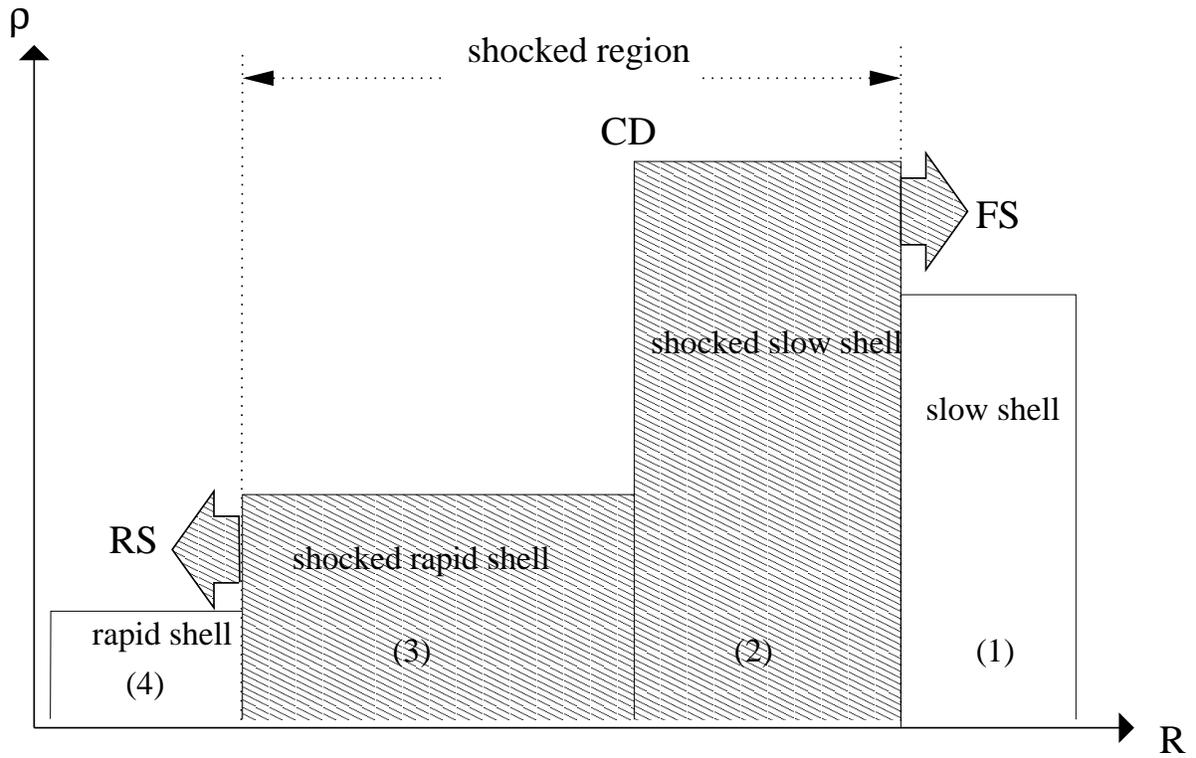}
\caption
%
{Sketch of a two-shell-collision where
a rapid shell catches up with a slower one at the CD frame.
Forward and reverse shocks (FS and RS) propagate from the contact discontinuity
(CD). We employ the conventional numbering for each region in the study
of GRB (e.g., Piran 1999). }\label{shock_is}
\end{figure}

In Fig. \ref{shock_is} 
we draw a schematic 
mass density profile during the shock propagation
in the interactions of rapid and slow shells.
Two shocks are formed: 
a reverse shock that goes 
into the rapid shell
and a forward shock that propagates into the slow shell. 
There are four regions: 
(1) the unshocked slow shell,
(2) the shocked slow shell,
(3) the shocked rapid shell, and
(4) the unshocked rapid shell.
Thermodynamic quantities, 
shch as 
rest mass density $\rho$, 
pressure $P$, 
and 
internal energy density $e$
are measured in the fluid rest frames.
We use the terminology 
of  {\it regions} $i$ ($i$=1, 2, 3, and 4) and
{\it position of discontinuity}
$j$ ($j$=FS, CD, and RS) where 
FS, CD, and RS stand for the forward shock front, 
contact discontinuity,
and reverse shock front, respectively.
The fluid velocity  
and Lorentz factor in the region $i$ 
measured in the ISM rest frame  
are expressed as
$v_{i}(=\beta_{i}c)$, and $\Gamma_{i}$, respectively.
The relative velocity 
and Lorentz factor of the fluids 
in the regions $i$ and $j$ are denoted as 
$v_{ij}(=-v_{ji}=\beta_{ij}c=-\beta_{ji}c)$ 
and $\Gamma_{ij}(=\Gamma_{ji})$, respectively. 
Throughout this work, we use the assumption of $\Gamma_{i}\gg 1$.

We first count the numbers of 
quantities and the shock jump conditions.
Each region is specified by three physical quantities;
rest-mass density $\rho_{i}$, 
pressure $P_{i}$, and
velocity $v_{i}$ measured in the ISM rest frame.
Forward and reverse shock speeds measured
in the frame of unshocked regions 
(i.e., regions 1 and 4, respectively) 
are two other quantities.
In all, there are $3\times 4+2=14$ physical quantities. 
The total number of the jump conditions 
at the forward shock (FS), 
reverse shock (RS), 
and contact discontinuity (CD) 
is $3+3+2=8$.
Hence, given $3+3=6$ upstream quantities 
for each shock, 
we can obtain the remaining $8$ downstream 
quantities by using $8$
jump conditions.

Following Blandford \& McKee (1976), we consider 
the limit of strong shock, 
$P_{2}/n_{2}\gg P_{1}/n_{1}$,
and adopt the assumption that the upstream matter is cold.
As an equation of state (EOS), we take
\begin{eqnarray}
P_{i}=(\hat{\gamma}_{i}-1)(e_{i}-\rho_{i})
\end{eqnarray}
where $\hat{\gamma}_{i}$ is an adiabatic index.
The jump conditions for the forward shock 
are written as follows:
\begin{eqnarray}\label{eq:FS}
\Gamma_{\rm FS1}^{2}
=\frac{(\Gamma_{\rm 12}+1)[\hat{\gamma}_{2}(\Gamma_{\rm 12}-1)+1]^{2}}
{\hat{\gamma}_{2}(2-\hat{\gamma}_{2})(\Gamma_{\rm 12}-1)+2} ,\nonumber \\
e_{2}=\Gamma_{\rm 12}\rho_{2} \, ,
\qquad
\frac{\rho_{2}}{\rho_{1}}=
\frac{\hat{\gamma}_{2}\Gamma_{12}+1}{\hat{\gamma}_{2}-1} \, ,
\end{eqnarray}
where $\Gamma_{12}=\Gamma_{1}\Gamma_{2}(1-\beta_{1}\beta_{2})$,
and
$\Gamma_{\rm FS1}$
is the Lorentz factor of forward shock measured 
in the rest frame of the unshocked slow shell. 
In the relativistic limit, the adiabatic index is ${\hat \gamma}_{2}=4/3$.
Using the same assumptions as in the forward shock,
the jump conditions for the reverse shock are given by:
\begin{eqnarray}\label{eq:RS}
\Gamma_{\rm RS4}^{2}
=\frac{(\Gamma_{\rm 34}+1)[\hat{\gamma}_{3}(\Gamma_{\rm 34}-1)+1]^{2}}
      {\hat{\gamma}_{3}(2-\hat{\gamma}_{3})(\Gamma_{\rm 34}-1)+2},\nonumber \\
e_{3}=\Gamma_{\rm 34}\rho_{3} \, ,
\qquad
\frac{\rho_{3}}{\rho_{4}}=
\frac{\hat{\gamma}_{3}\Gamma_{34}+1}{\hat{\gamma}_{3}-1} \, ,
\end{eqnarray}
where $\Gamma_{34}=\Gamma_{3}\Gamma_{4}(1-\beta_{3}\beta_{4})$,
and $\Gamma_{\rm RS4}$
is the Lorentz factor of the reverse shock measured in 
the rest frame of the unshocked rapid shell. 
The equality of pressures and velocities across the contact discontinuity 
gives
\begin{eqnarray}
P_{2}=P_{3} ,
\qquad
\Gamma_{2}=\Gamma_{3} \, .
\end{eqnarray}
Before a shock breakout,
$\Gamma_{2}=\Gamma_{3}=\Gamma_{\rm CD}$ is satisfied.
It may be useful to introduce the ratio 
$f\equiv\rho_{4}/\rho_{1}$, which
can be obtained from $P_{2}=P_{3}$, as
\begin{eqnarray}\label{fblazar}
f\equiv\frac{\rho_{4}}{\rho_{1}}
=\frac{(\hat{\gamma}_{2}\Gamma_{12}+1)(\Gamma_{12}-1)}
{(\hat{\gamma}_{3}\Gamma_{34}+1)(\Gamma_{34}-1)} \ .
\end{eqnarray}

Throughout this paper, we set $P_{1}=P_{4}=0$. 
Then, we take $\rho_{1}$, $\rho_{4}$, 
$\Gamma_{1}$, and $\Gamma_{4}$ as ramaining $4$ upstream parameters.
Then, with $8$ shock conditions given above,
we can obtain $8$ downstream quantities 
$\rho_{2}$
$e_{2}$, 
$v_{2}$,
$v_{\rm FS}$,
$\rho_{3}$, 
$e_{3}$,
$v_{3}$, and
$v_{\rm RS}$.

\subsection{Time Scales of Wave Propagations}

Here we evaluate seven time scales of
relevance when
the shock and rarefaction waves cross the 
colliding shells.
They are
useful to understand the hydrodynamical evolution
of two-shell-collisions.
We measure these time scales in the rest frame of CD
(hereafter we call it the CD frame)
because it facilitates the comparison with each other.
In contrast, most of the previous papers
used the ISM frame in measuring
crossing time of shock 
(e.g., Sari \& Piran 1995; Panaitescu et al. 1997). 
Here we need to introduce 
new physical
paramaters, 
the shell widths measured 
in the ISM frame, 
$\Delta_{\rm r}$, and
$\Delta_{\rm s}$,
where subscripts $r$ and $s$ 
represent rapid and slow shells, respectively.
In the ISM frame, 
the upstream parameters are as follows:
Lorentz factors
$\Gamma_{\rm r}(=\Gamma_{4})$, and
$\Gamma_{\rm s}(=\Gamma_{1})$,
rest mass densities
$\rho_{\rm r}$, and 
$\rho_{\rm s}$.
With $2+4=6$ physical parameters, 
we can uniquely specify 
the initial condition.
Note that the regions 1 and 4 disappear
after FS and RS break out, respectively.

In the CD frame, we rewrite them as follows;
$\Delta_{\rm r}^{'}$,
$\Delta_{\rm s}^{'}$,
$\Gamma_{\rm r}^{'}(=\Gamma_{4}^{'}=\Gamma_{34})$, and
$\Gamma_{\rm s}^{'}(=\Gamma_{1}^{'}=\Gamma_{12})$
during the shock propagation in the shells.
After the shock breaks out of the shell,
the velocity is not uniform and determined
by the propagation of rarefaction wave.
Note that once  
we choose the CD frame, 
$\Gamma_{4}^{'}$ and $\Gamma_{1}^{'}$ 
are not independent of each other 
(see e.g., Eq. (\ref{eq:g4'}) below). 

The time in which FS crosses the slow shell, 
$t_{\rm FS}^{'}$, is given by
\begin{eqnarray}\label{eq:tFS'}
t_{\rm FS}^{'}&=& 
\frac{\Delta_{\rm s}^{'}}
{|\beta_{1}^{'}|
+|\beta_{\rm FS}^{'}|}
\nonumber\\
&=&\Delta_{\rm s}^{'}
\left[|\beta_{1}^{'}|+
\left(1-
\frac{1}
{\Gamma_{1}^{'2}\Gamma_{\rm FS1}^{2}
(1-
|\beta_{1}^{'}|
|\beta_{\rm FS1}|)^{2}}
\right)^{1/2}
\right]^{-1} ,
\end{eqnarray}
where we use Eq. (\ref{eq:FS})
and $\beta_{\rm FS}^{'2}=1-1/\Gamma_{\rm FS}^{'2}$.
Thus, we can express $t_{\rm FS}^{'}$ as 
a function of model parameters
and $\Gamma_{1}^{'}$ which is also given implicitly from
the model parameters.
Similarly, the RS crossing-time 
in the rapid shell, $t_{\rm RS}^{'}$, is 
\begin{eqnarray}\label{eq:tRS'}
t_{\rm RS}^{'}&=& 
\frac{\Delta_{\rm r}^{'}}
{|\beta_{4}^{'}|
+
|\beta_{\rm RS}^{'}|}
\nonumber\\
&=&\Delta_{\rm r}^{'}
\left[|\beta_{4}^{'}|+
\left(1-
\frac{1}
{\Gamma_{4}^{'2}\Gamma_{\rm RS4}^{2}
(1-|\beta_{4}^{'}|
|\beta_{\rm RS4}|)
^{2}}
\right)^{1/2}
\right]^{-1} ,
\end{eqnarray}
where we use Eq. (\ref{eq:RS}) and 
$\beta_{\rm RS}^{'2}=1-1/\Gamma_{\rm RS}^{'2}$.
It is important to note 
that $\Gamma_{1}^{'}$ and $\Gamma_{4}^{'}$ are not independent
but related by the equation below,
\begin{eqnarray}\label{eq:g4'}
\Gamma_{4}^{'}= 
\frac{-f(1-\hat{\gamma}_{3})+
\sqrt{
f^{2}(1-\hat{\gamma}_{3})^{2}
+4f\hat{\gamma}_{3}
\left[
(\Gamma_{1}^{'}-1)(\hat{\gamma}_{2}\Gamma_{1}^{'}+1)+f
\right]
}}
{2f\hat{\gamma}_{3}}.
\end{eqnarray}
It is expected that
after FS has crossed the slow  shell,
a rarefaction wave (hereafter we call it FR)
propagates into the shocked slow shell 
(e.g., Panaitescu et al. 1997).
The  sound speed is given by (e.g., Mihalas \& Mihalas 1984)
\begin{eqnarray}
c_{\rm s}^{2}=
\left(\frac{\partial P}{\partial e}\right)_{\rm ad}
=\frac{\hat{\gamma} P}{ e+P} .
\end{eqnarray}
Thus,
the time at which FR reaches CD, 
$t_{\rm FR-CD}^{'}$, is given by 
\begin{eqnarray}\label{eq:tFRCD'}
t_{\rm FR-CD}^{'}&=& t_{\rm FS}^{'}
+\frac{\Delta_{\rm s,FS}^{'}}
      {c_{\rm s2}}   \nonumber \\ 
&=&t_{\rm FS}^{'}+
\Delta_{\rm s}^{'}
\Gamma_{1}^{'}
\frac{\hat{\gamma}_{2}-1}{\hat{\gamma}_{2}\Gamma_{1}^{'}+1}
\left[\frac{
\hat{\gamma}_{2}
(\hat{\gamma}_{2}-1)
(\Gamma_{1}^{'}-1)
}{\hat{\gamma}_{2}\Gamma_{1}^{'}-\hat{\gamma}_{2}+1}c^{2}
\right]^{-1/2}  \ ,
\end{eqnarray}
where
$\Delta_{\rm s,FS}^{'}$ is the width of the slow shell 
just after FS reaches the end of the shell.
This is obtained by the mass conservation (e.g., Spada et al. 2001) 
where $(\hat{\gamma}_{2}-1)/(\hat{\gamma}_{2}\Gamma_{1}^{'}+1)$
is the compression factor 
of the slow shell and $\Gamma_{1}^{'}$
is the factor from the Lorentz contraction.
Just the same way,
the corresponding time, $t_{\rm RR-CD}^{'}$, 
at which the rarefaction wave (hereafter we call it RR)
generated at the 
RS breakout reaches CD is given by
\begin{eqnarray}\label{eq:tRRCD'}
t_{\rm RR-CD}^{'}&=& t_{\rm RS}^{'}
+\frac{\Delta_{\rm r,RS}^{'}}
      {c_{\rm s3}}  \nonumber \\ 
&=&t_{\rm RS}^{'}+
\Delta_{\rm r}^{'}
\Gamma_{4}^{'}
\frac{\hat{\gamma}_{3}-1}{\hat{\gamma}_{3}\Gamma_{4}^{'}+1}
\left[\frac{
\hat{\gamma}_{3}
(\hat{\gamma}_{3}-1)
(\Gamma_{4}^{'}-1)
}{\hat{\gamma}_{3}\Gamma_{4}^{'}-\hat{\gamma}_{3}+1}c^{2}
\right]^{-1/2} .
\end{eqnarray}
In the case of $t_{\rm RR-CD}^{'}>t_{\rm FR-CD}^{'}$,
only $t_{\rm FR-CD}^{'}$ is an actual time and 
$t_{\rm RR-CD}^{'}$ is a virtual time 
which does not exist in reality. 
The opposite case is also true.

In the case of $t_{\rm RR-CD}^{'} < t_{\rm FR-CD}^{'}$,
we have the time at which two rarefaction waves collide each other,
$t_{\rm RR-FR}^{'}$,
as 
\begin{eqnarray}\label{eq:tRRFR'}
t_{\rm RR-FR}^{'}
&\sim& t_{\rm RR-CD}^{'}
+\frac{\Delta_{\rm s,RR-CD}^{'}}
      {2c_{\rm s2}}\nonumber \\ 
&\sim&
t_{\rm RR-CD}^{'}+
\frac{\Delta_{\rm s}^{'}\Gamma_{1}^{'}}{2c_{\rm s2}}
\frac{\hat{\gamma}_{2}-1}
{\hat{\gamma}_{2}\Gamma_{1}^{'}+1}
\left(
\frac{t_{\rm FR-CD}^{'}-t_{\rm RR-CD}^{'}}
{t_{\rm FR-CD}^{'}-t_{\rm FS}^{'}}
\right) ,
\end{eqnarray}
where $\Delta_{\rm s,RR-CD}^{'}$ is the width of 
the part of slow shell which FR has not passed through 
yet at $t_{\rm RR-CD}^{'}$.
Since both RR and FR propagate at 
the speed $c_{\rm s2}$ after  
$t_{\rm RR-CD}^{'}$, 
the above equation includes factor 2. 
Note that after the rarefaction wave crosses CD,
the pressure gradient appears at CD. As a result,
the CD begins to move from $x^{'}=0$ in the CD frame.
Thus,
Eqs. 
(\ref{eq:tRRFR'}),
(\ref{eq:tRRFR'2}),
(\ref{eq:tRRFS'}), and 
(\ref{eq:tFRRS'})  are apploximated estimations.
Similarly, 
in the case of $t_{\rm FR-CD}^{'} < t_{\rm RR-CD}^{'}$,
we have
\begin{eqnarray}\label{eq:tRRFR'2}
t_{\rm RR-FR}^{'}
&\sim& t_{\rm FR-CD}^{'}
+\frac{\Delta_{\rm r,FR-CD}^{'}}
      {2c_{\rm s3}}\nonumber \\ 
&\sim&
t_{\rm FR-CD}^{'}+
\frac{\Delta_{\rm r}^{'}\Gamma_{4}^{'}}{2c_{\rm s3}}
\frac{\hat{\gamma}_{3}-1}
{\hat{\gamma}_{3}\Gamma_{4}^{'}+1}
\left(
\frac{t_{\rm RR-CD}^{'}-t_{\rm FR-CD}^{'}}
{t_{\rm RR-CD}^{'}-t_{\rm RS}^{'}} 
\right) .
\end{eqnarray}

The time scale in which RR catches up with
the propagating forward shock (FS),
$t_{\rm RR-FS}^{'}$,
can be estimated as
\begin{eqnarray}\label{eq:tRRFS'}
t_{\rm RR-FS}^{'}\sim
\frac{c_{\rm s,2}t_{\rm RR-CD}^{'}}
      {c_{\rm s2}-\beta_{\rm FS}^{'}c} \ .
\end{eqnarray}
Similarly,
the time-scale in which FR catches up with 
the reverse shock (RS) is approximated as 
\begin{eqnarray}\label{eq:tFRRS'}
t_{\rm FR-RS}^{'}\sim
\frac{c_{\rm r,3}t_{\rm FR-CD}^{'}}
      {c_{\rm s3}-\beta_{\rm RS}^{'}c} \ .
\end{eqnarray}

\subsection{Numerical Simulation}

We complementarily
perform the special 
relativistic hydrodynamical simulations.
The detail of the code is given in
Mizuta et al. (2004).
To sum up,
the code is based on an approximate relativistic
Riemann solver.
The numerical flux is derived from Marquina's flux formula
(Donat \& Marquina 1996).
This code is originally
second order in space using the so-called MUSCL method.
In this study, however,
this is slightly compromised 
for numerical stability. 
We assume plane symmetry and treat 
one dimensional motions of shells.
In discussing the propagation of shock and rarefaction waves, 
we choose the CD frame.
Given the ratio $\Gamma_{\rm r}/\Gamma_{\rm s}$ 
in the ISM frame and the value of $\Gamma_{\rm s}$,
we can determine 
the Lorentz transformation to the CD frame easily
because the CD Lorentz factor 
$\Gamma_{\rm CD}(=\Gamma_{2}=\Gamma_{3})$
measured in the  ISM frame can be derived by solving 
Eq. (\ref{fblazar}).
We should note that
the number of free parameters is reduced from 6 to 5
because we have already fixed the frame.
As for the EOS,
we assume for simplicity that
$\hat{\gamma}_{3}=\hat{\gamma}_{2}=4/3$
for $\Gamma_{34}>2$, and
$\hat{\gamma}_{3}=\hat{\gamma}_{2}=5/3$
otherwise.
Although this simplification gives 
slightly inaccurate estimation
on the speeds of wave propagations, 
there is little effect on our conclusions
in this work.

We start the calculation at $t=0$ when the collision of
two shells has just begun.
Thoughout this paper,
we set $\Delta_{s}^{'}/c=1$ 
and  $\rho_{r}=1$ 
as units in numerical simulations.
Initially, two shells have opposite velocities,
namely, $v_{\rm r}^{'} > 0$ and $v_{\rm s}^{'} < 0$.
In the previous section, we did not 
impose any conditions for the plasma
surrounding the two shells.
We only assumed that the boundary of
each shell will be kept intact
during the passage of 
shocks and rarefaction waves.
For our numerical runs,
we put plasma of low rest mass density 
$(10^{-4} \ll 1,\rho_{\rm s}/\rho_{\rm r})$
outside of the two shells. They have 
the same velocity
and pressure as the adjacent shell.
At first,
the boundary condition at the left boundary is
a steady inflow of dilute plasma.
When the reverse shock or the rarefaction 
wave reaches the rapid shell's boundary,
the velocity of the dilute plasma is set to be zero instantaneously
to reduce the effect of the interaction between the shock and
the dilute plasma.
At the same time, the left boundary condition is set to be a
free outflow.
The treatment of the right side dilute plasma and 
the boundary condition is
the same as that of the left side.

\section{SHELL DYNAMICS AFTER COLLISION}

\subsection{Shell Splitting}
\subsubsection{General Consideration}

Here we classify the types of 
the mass density profile in the merged shell
based on the order of the times obtained in the 
previous section.
Table 1 gives the complete set of the possible
orders.
Although
there are various cases in the orders,
the density profile, in particular, the
splitting feature is governed  by
 the two criteria as follows.
\begin{enumerate}
\item[(I)]

When $t_{\rm RR-CD}^{'}<t_{\rm FR-CD}^{'}$ for
$\rho_{\rm r}>\rho_{\rm s}$
or
$t_{\rm FR-CD}^{'}<t_{\rm RR-CD}^{'}$
for
$\rho_{\rm s}>\rho_{\rm r}$
is satisfied, 
the splitting occurs at the CD
since
the rarefaction wave going
from the larger density region (region 2)
into the
smaller density region (region 3) 
makes a dip in the latter region.

\item[(II)]

When a pair of rarefaction waves propagating in 
the opposite directions collide with each other,
the density begins to decrease at the collision point
and the splitting feature emerges.
Hence the existence of
$t_{\rm RR-FR}$ implies splitting feature. 

\end{enumerate}

\begin{figure}
\plotone{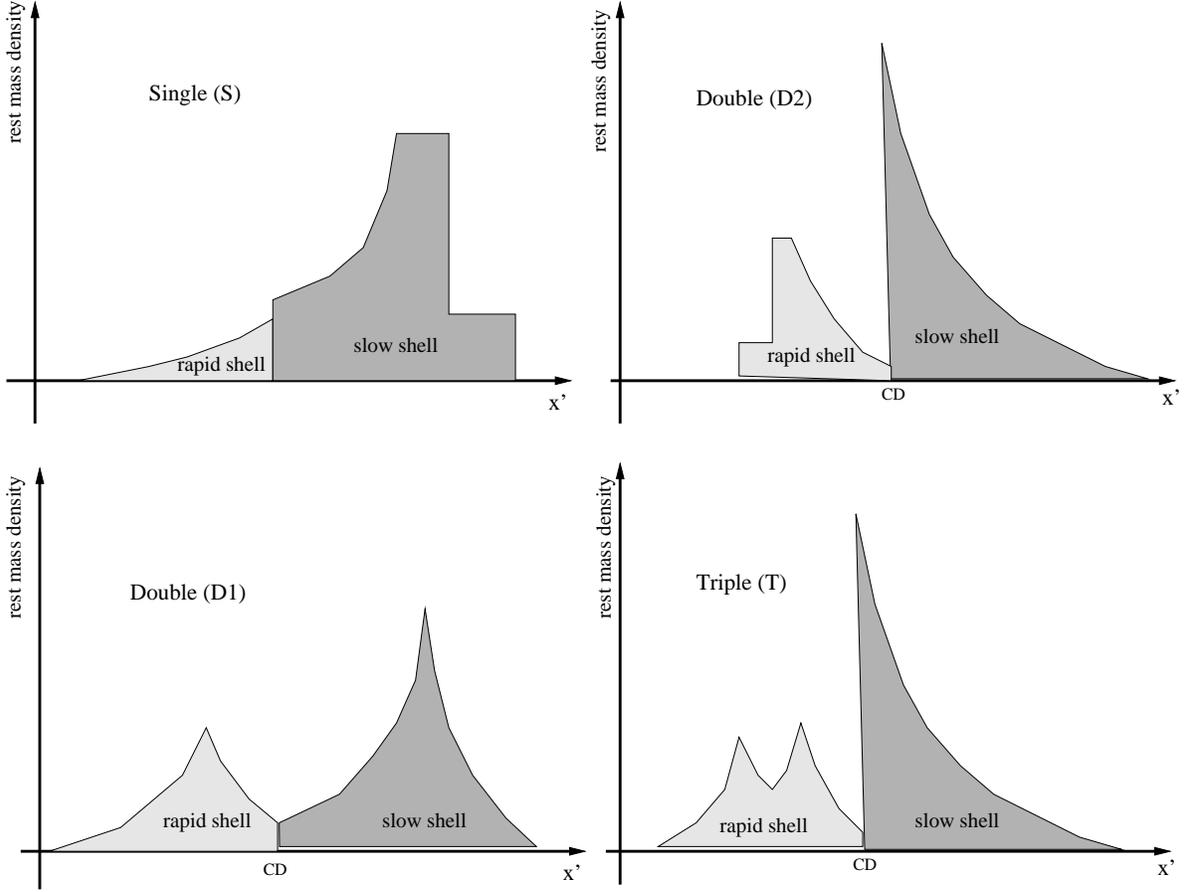}
\caption
%
{Schematic snapshot of the rest mass density of 
two shells. Here we assume $\rho_{\rm s}>\rho_{\rm r}$. 
The corresponding time scale relation in each case is
shown in Table 1.
Picture S shows the case where 
RR catches up with the propagating FS and
has a single-peak.
Picture D1 shows the case where 
both of FS and RS
cross the shells and the 
rarefaction waves propagating from both sides 
dig a dip and give a double-peaked profile.
Picture D2 shows the case where 
FR reaches  CD and dig a dip. 
Hence the profile is double-peaked.
Picture T shows the case of the combination of
D1 and D2.
Then the final profile is triple-peaked.}\label{rhocst}
\end{figure}

\begin{figure}
\plotone{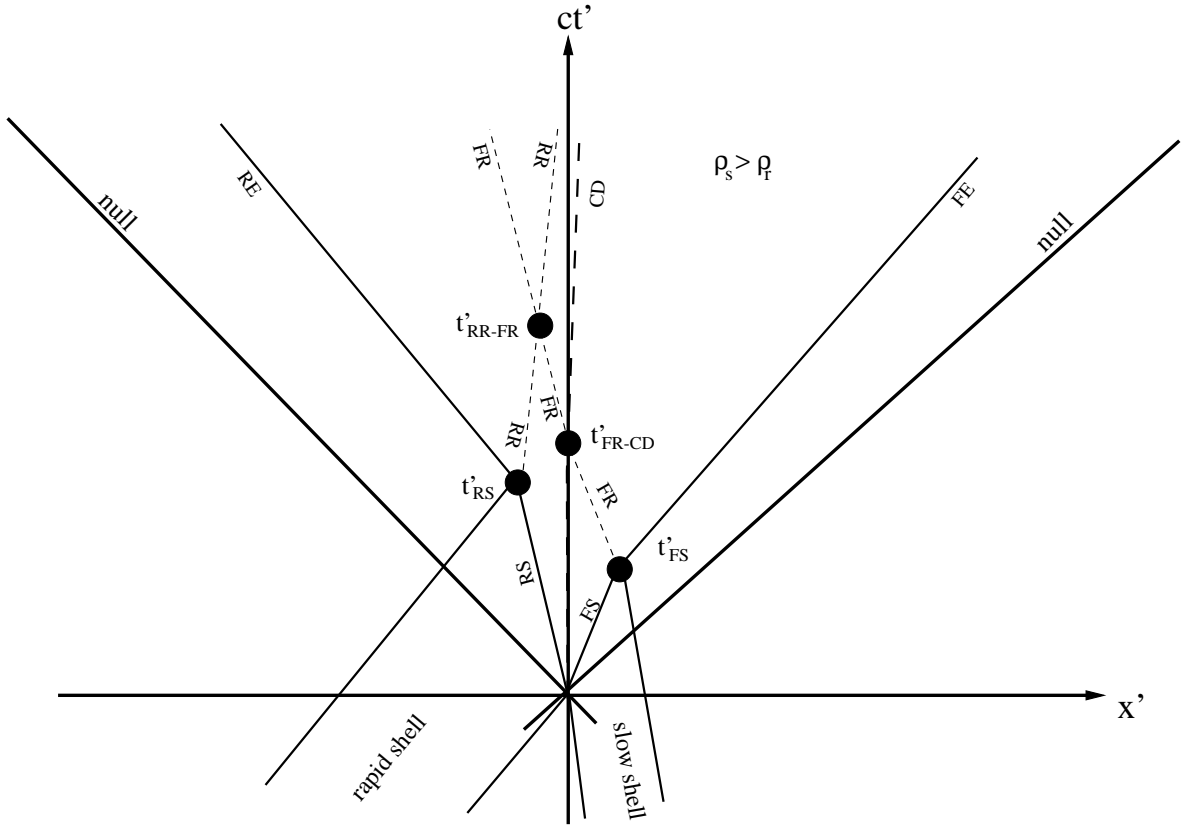}
\caption
%
{Schematic picture
of space-time diagram of a two-shell-interaction in the
CD frame.
Here we assume $\rho_{\rm s}>\rho_{\rm r}$. 
A forward shock (FS) and a reverse shock (RS) run through the 
slow and rapid shells, respectively.
After the shocks break out, the rarefaction waves 
propagate into the shells (FR and RR) 
and the shells spread (FE and RE).
This corresponds to case 7 in Table 1.}\label{cst}
\end{figure}

Based on these two criteria, 
the mass density profile is classified 
into four types 
and shown in Fig. \ref{rhocst}.
If both criteria are satisfied,
then the mass density has triple peaks.
We show the corresponding 
schematic picture of space time diagram in Fig. \ref{cst}.
If only one criteria is met,
then the double-peaked profile is realized.
When neither condition is satisfied,
the single peak is obtained.

\subsubsection{GRBs and Blazars}

Here we apply the above general consideration to the specific cases and
examine
which kind of  rest mass density profile is realized
in GRBs and blazars.
We assume that the widths of two shells are
same in the ISM frame
which is written as $\Delta_{\rm r}/\Delta_{\rm s}=1$ 
(see, e.g., KS01).
We consider following three cases
since it seeems natural 
to suppose that ejected shells 
from the central engine have a correlation among them;
(1) the energy of bulk motion of the rapid shell ($E=\Gamma mc^{2}$) 
equals to that of the slow one in the ISM frame
(we refer to it as ``equal energy (or $E$)''),
(2) the mass of rapid shell ($m=\rho \Gamma\Delta$)
equals to that of the slow one 
(hereafter we call it ``equal mass (or $m$)''),
and 
(3) the rest mass density of rapid shell equals to that of the slow one 
(hereafter we call it ``equal rest mass density (or $\rho$)''),
\begin{eqnarray}
\rho_{\rm r}
&=&
\rho_{\rm s}
\qquad   ({\rm equal}  \ \rho) ,  \nonumber \\
\rho_{\rm r}
\Delta_{\rm r}\Gamma_{\rm r}
&=&
\rho_{\rm s}
\Delta_{\rm s}
\Gamma_{\rm s}
\qquad  ({\rm equal}  \ m) , \nonumber \\
\rho_{\rm r}
\Delta_{\rm r}
\Gamma_{\rm r}^{2}c^{2}
&=&
\rho_{\rm s}
\Delta_{\rm s}
\Gamma_{\rm s}^{2}c^{2}
\qquad  ({\rm equal} \ E) .
\end{eqnarray}
Note that in the case of
$\Delta_{\rm r}=\Delta_{\rm s}$
and $\Gamma_{\rm r} >\Gamma_{\rm s}$,
$\rho_{\rm s}$ is always
larger than  $\rho_{\rm r}$.
This leads to the absence of 
$t_{\rm RR-FS}$. 
For all cases, we have
$3+1+1=5$ parameters. 
We take the ratio 
$\Gamma_{\rm r}/\Gamma_{\rm s}$ 
as the last parameter 
and vary its value.
This completes the six model parameters.

\begin{figure}
\plotone{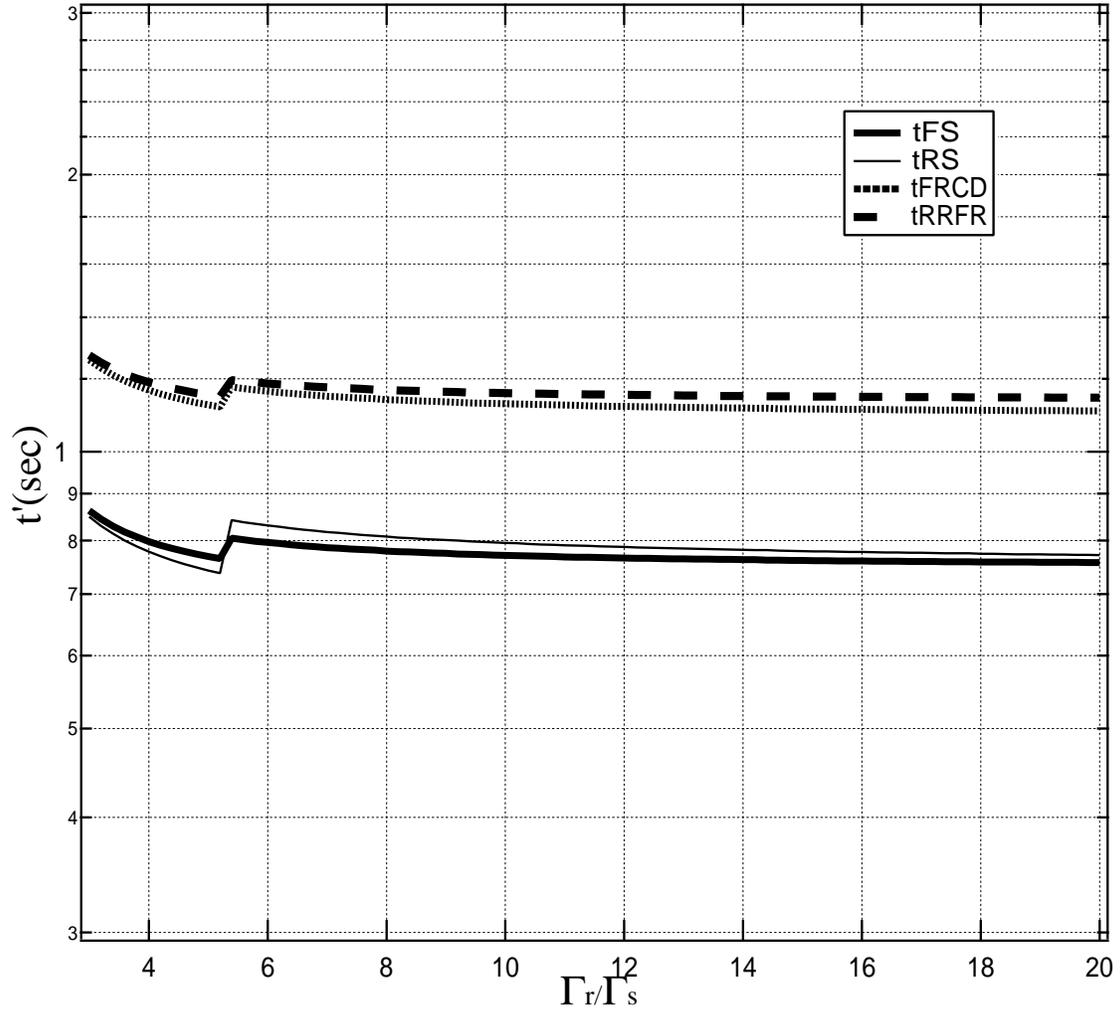}
\caption
%
{$\Gamma_{\rm r}/\Gamma_{\rm s}$ dependence of the 
various time scales for the ``equal E''case.
In the whole range, case 4 (Triple) is realized.
The slight jumps of the time scales 
at $\Gamma_{\rm r}/\Gamma_{\rm s}\sim 5$
correspond to the abrupt change of adiabatic index.
The softening of EOS in the relativistic
regime gives slower shock waves
and the time scales get longer accordingly.
}\label{GRB}
\end{figure}

\begin{figure}
\plottwo{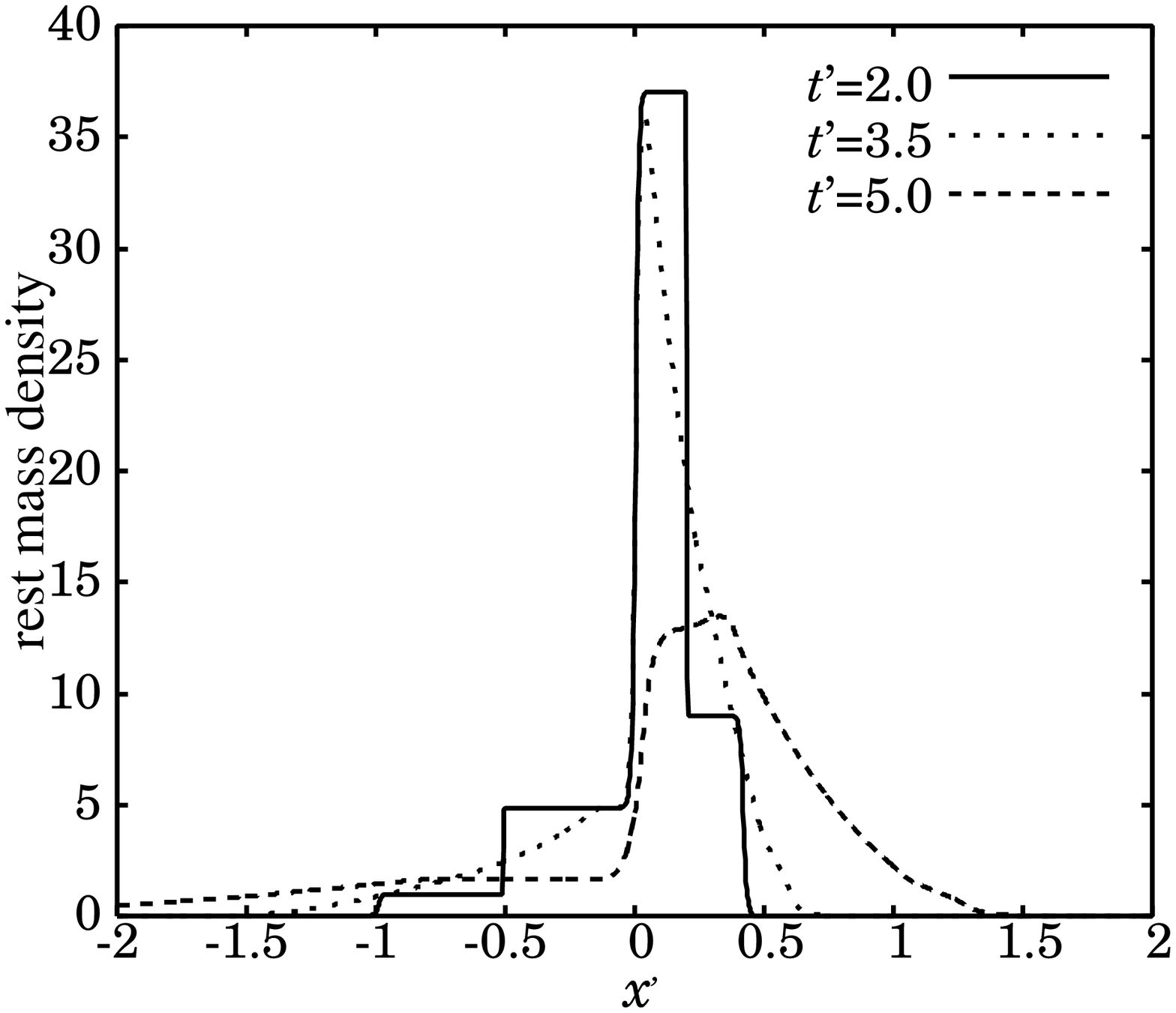}{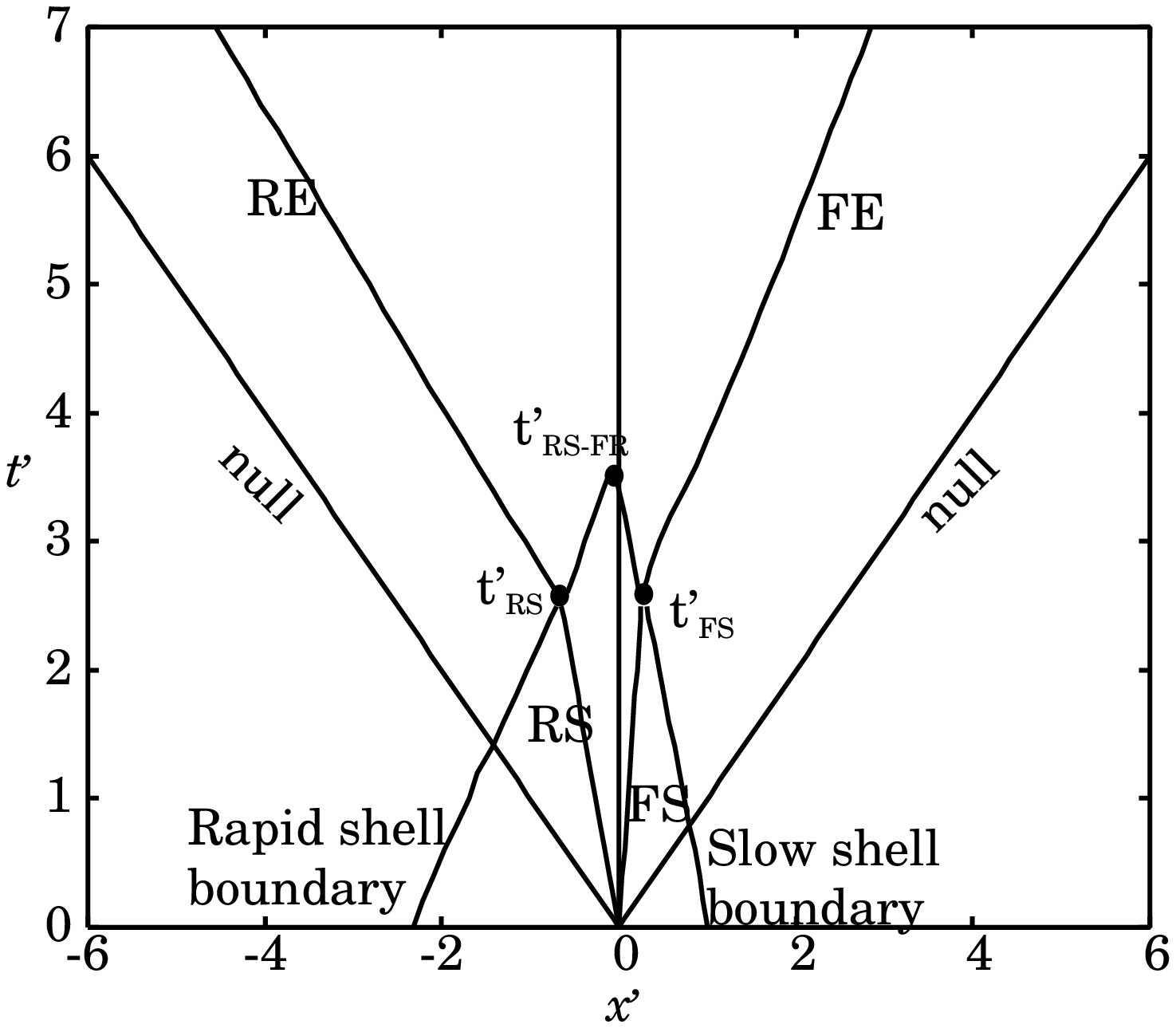}
\caption
%
{
Left:
Time evolution of 
the rest mass density profile in the CD frame
in the case of ``equal $E$''. 
The parameter is chosen
so that $\Gamma_{\rm r}/\Gamma_{\rm s}=3$ in the ISM frame.
The parameters in the CD frame are shown in Table 2.
Throughout our numerical simulations, we set 
$\Delta_{\rm s}^{'}/c=1$ and
$\rho_{\rm s}=1$
as units.
Right:
Space-time diagram of shock and rarefaction waves propagations.
As is shown in the text, $t_{\rm RS}\sim t_{\rm FS}\sim 3$.
RE spreads at the speed $\sim 0.9c$
while FE spreads at the speed $\sim 0.6c$.}
\label{fig:e31}
\end{figure}

\begin{figure}
\plottwo{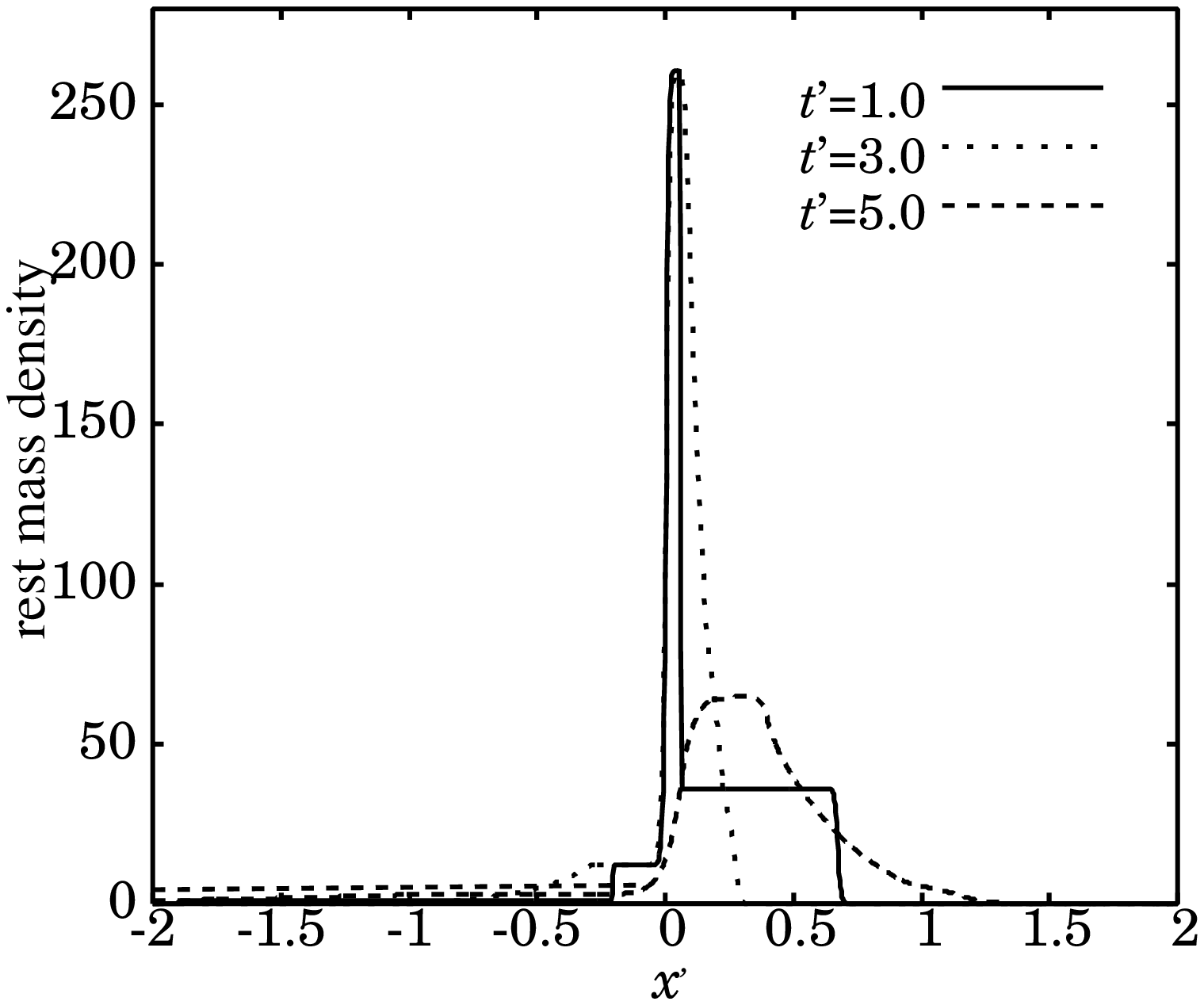}{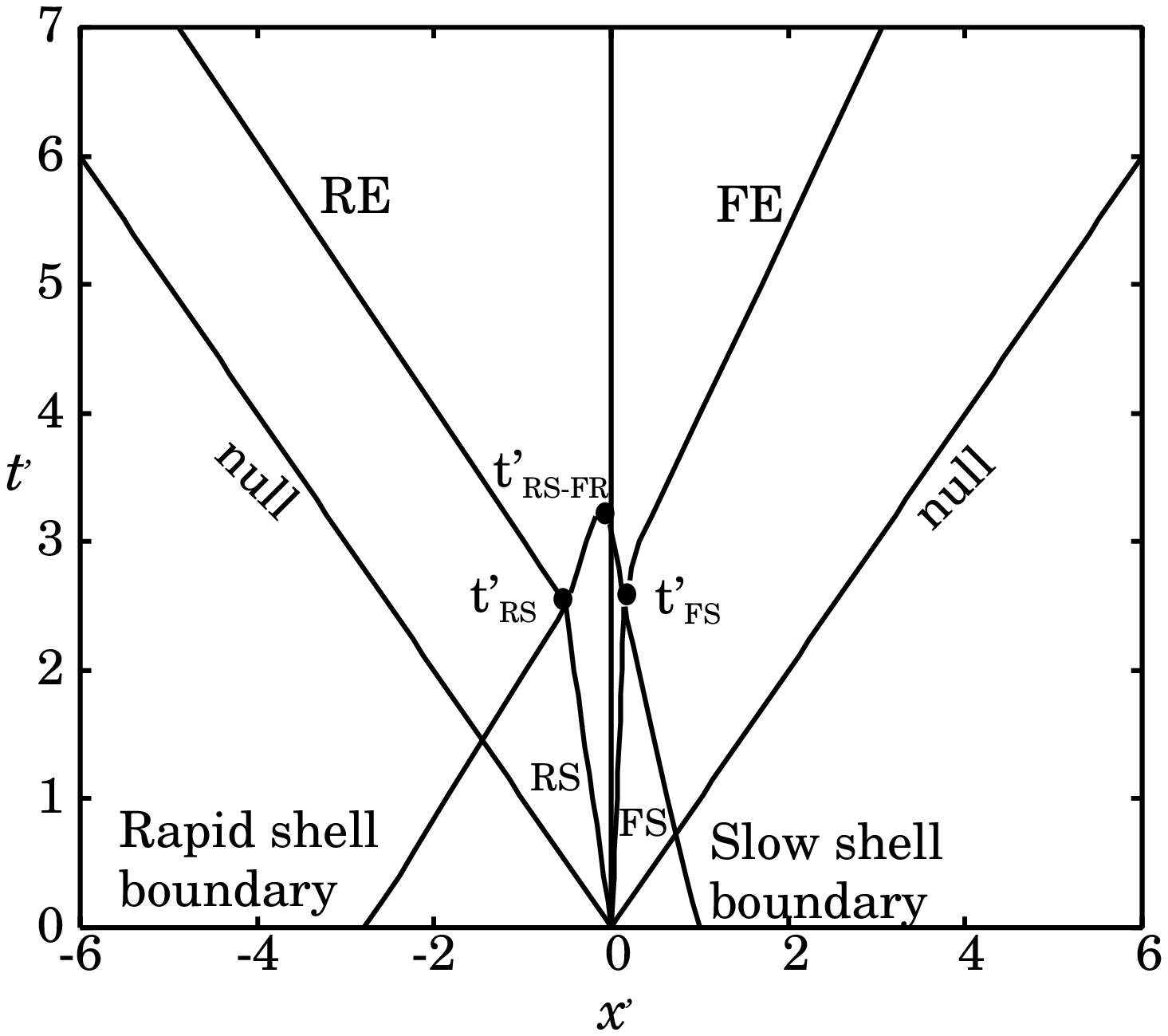}
\caption
{
Left:
Time evolution of the 
rest mass density profile in the CD frame
for ``equal $E$''.
In the ISM frame,
$\Gamma_{\rm r}/\Gamma_{\rm s}=6$. 
The parameters are given in Table 2.
Right:
Space-time diagram of shock and rarefaction waves propagations.
Both FS and RS propagate faster than those for
$\Gamma_{\rm r}/\Gamma_{\rm s}=3$ and
$t_{\rm FS}\sim t_{\rm RS}\sim 3$.
RE spreads at the speed $\sim c$
while FE spreads at the speed $\sim 0.7c$.}
\label{fig:e61}
\end{figure}

The various time scales 
for the ``equal $E$'' case in GRBs
are shown as a function of 
$\Gamma_{\rm r}/\Gamma_{\rm s}$ 
in Fig. \ref{GRB}.
Here we set
$\Gamma_{\rm s}=10^{2}$,
$\Delta_{\rm s}^{'}=10^{10}$ cm, and
$\rho_{\rm s}=10^{-10}$  g cm$^{-3}$
in the slow shell as an example.
Slight jumps of time scales
are seen
at $\Gamma_{\rm r}/\Gamma_{\rm s}\sim 5$ in the Figure.
They are caused  
by the abrupt change of adiabatic index between
the non-relativistic regime and relativistic one.
The softening of EOS in the relativistic regime
leads to slower shock waves propagation in the CD frame.
In the whole range,
the criteria (I) and (II)
given in the previous section are both satisfied. 
Therefore, the triple-peaked profile 
is expected (No. 4 in Table 1) in principle.
However, 
the criterion (I) is satisfied only marginally. 
As a result, two peaks are not remarkable.
It is worthwhile to obtain order estimations of 
$\Delta_{\rm r}^{'}/\Delta_{\rm s}^{'}$ and 
$|\beta_{\rm r}^{'}/\beta_{\rm s}^{'}|$
by using a simple apploximation of 
$\Gamma_{\rm CD}\sim\Gamma_{\rm m}$
($\Gamma_{\rm m}$ is the Lorentz factor of
marged shell obtained by two-mass-collision model
and it is given in the next subsection)
in spite of some discrepancy
with the exact solution 
of Eq. (\ref{fblazar}).
We have
\begin{eqnarray}
\frac{\Delta_{\rm r}^{'}}{\Delta_{\rm s}^{'}}
=
\frac{\Delta_{\rm r}\Gamma_{\rm r}\Gamma_{\rm s}^{'}}
{\Delta_{\rm s}\Gamma_{\rm s}\Gamma_{\rm r}^{'}}
\simeq 
\left(
\frac{\Gamma_{4}}{\Gamma_{1}}
\right)^{2}
\left(
\frac{\Gamma_{1}^{2}+\Gamma_{2}^{2}}{\Gamma_{3}^{2}+\Gamma_{4}^{2}}
\right),
\qquad
\left|\frac{\beta_{\rm r}^{'}}{\beta_{\rm s}^{'}}\right|
\simeq 
\frac
{ (\Gamma_{1}^{2}+\Gamma_{2}^{2})
(-\Gamma_{3}^{2}+\Gamma_{4}^{2})}
{(-\Gamma_{1}^{2}+\Gamma_{2}^{2})
 (\Gamma_{3}^{2}+\Gamma_{4}^{2})}  .
\end{eqnarray}
%
As $\Gamma_{\rm r}/\Gamma_{\rm s}$ increases,
the ratios of shell widths 
and velocities in the CD frame 
go asymptotically to 
\begin{eqnarray}\label{eq:eqEwb}
\frac{\Delta_{\rm r}^{'}}{\Delta_{\rm s}^{'}}
\sim 3,
\qquad
\left|\frac{\beta_{\rm r}^{'}}{\beta_{\rm s}^{'}}\right|
\sim 3   .
\end{eqnarray}
This equation explains well the fact
that each time scale in Fig. \ref{GRB} has a weak dependence
on $\Gamma_{r}/\Gamma_{s}$. 
This is the reason why $t_{\rm RS}^{'}$ and
$t_{\rm FS}^{'}$ are very close to each other.
$t_{\rm RR-CD}^{'}$ is also close to
$t_{\rm FR-CD}^{'}$ simply because the sound speeds
in the both shocked regions are about a few ten percents of 
the light speed and close to each other.
The corresponding numerical results
are shown in Figs. 
\ref{fig:e31} and \ref{fig:e61}.
In these calculations,
we take
the cases of
$\Gamma_{\rm r}/\Gamma_{\rm s}=3$ and
$\Gamma_{\rm r}/\Gamma_{\rm s}=6$, respectively.
This implies a large density contrast
of 
$\rho_{\rm s}/\rho_{\rm r}=9$ and
$\rho_{\rm s}/\rho_{\rm r}=36$, respectively (see Table 2).
The collision of the rarefaction waves occurs in
the region with much lower density 
compared with region 2. 
As a result,  the peak of the profile
is smoothed out. 
For larger values of
$\Gamma_{\rm r}/\Gamma_{\rm s}$,
the density contrast between regions 2 and 3 becomes
clearer.
Hence we conclude that the ``equal energy'' case
essencially evolves into single-peaked profiles. 
The space-time diagram
obtained by the numerical simulation for ``equal $E$'' is 
shown in 
Figs.
\ref{fig:e31} and
\ref{fig:e61}.
From Fig. \ref{fig:e31}, we see 
that $t_{\rm FS}^{'}
\sim t_{\rm RS}^{'}\sim 3$ as is shown in
Eq. (\ref{eq:eqEwb}). In Fig. \ref{fig:e61},
these time scales become
close to $3$.

\begin{figure}
\plotone{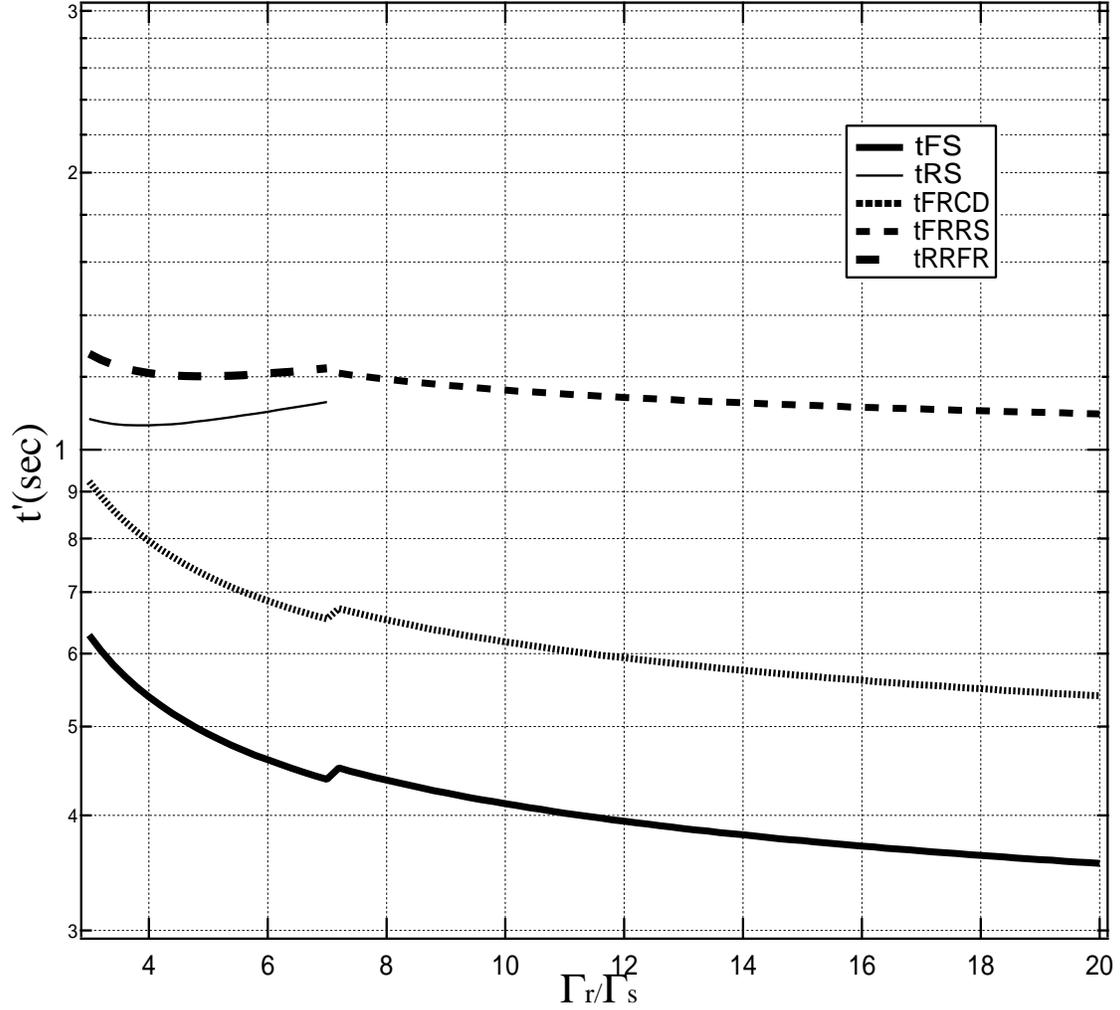}
\caption
%
{Time scales for ``equal mass''. 
When
$\Gamma_{\rm r}/\Gamma_{\rm s}$ is smaller than $\sim7$,
case 6 (Triple) is realized.
For 
$\Gamma_{\rm r}/\Gamma_{\rm s}$ larger than $\sim7$,
case 5 (D2) is realized.
The slight jumps of the time scales 
at $\Gamma_{\rm r}/\Gamma_{\rm s}\sim 7$
correspond to the abrupt change of adiabatic index.
The softening of EOS in the relativistic
regime gives slower shock waves
and the time scales get longer accordingly.}
\label{GRB_mass}
\end{figure}

\begin{figure}
\plottwo{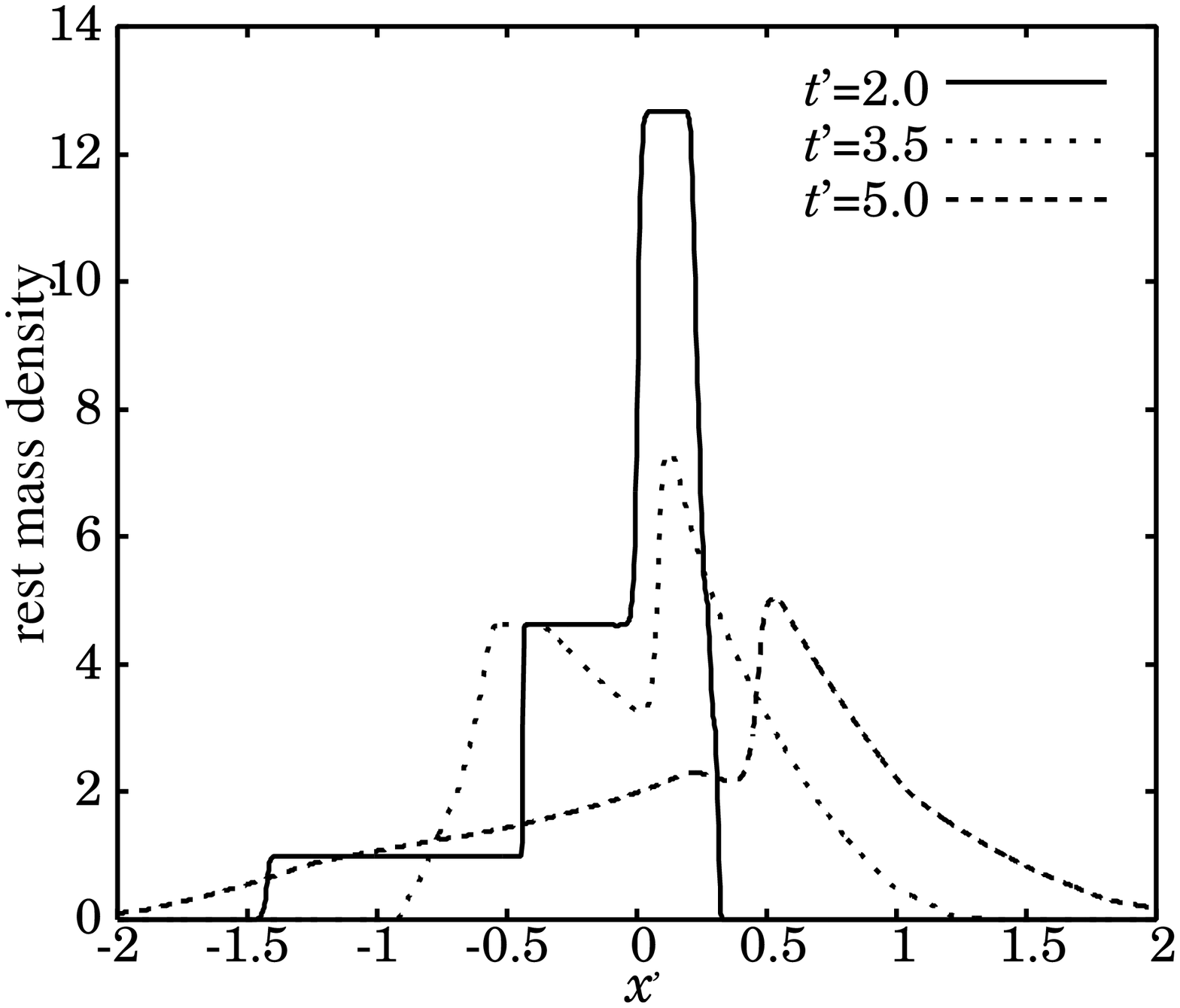}{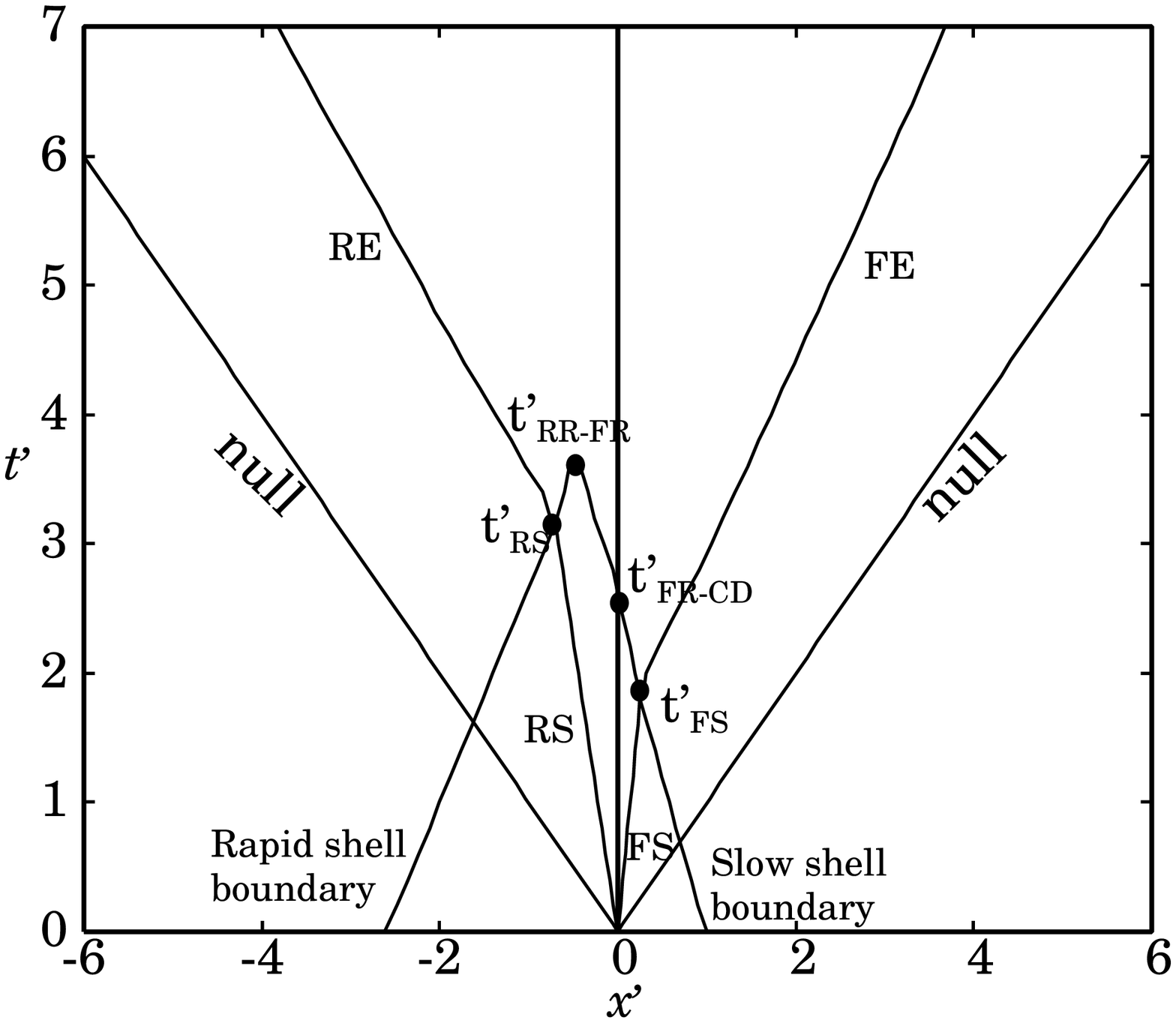}
\caption
%
{
Left:
Time evolution of 
the rest mass density profile in the CD frame for ``equal $m$''. 
In the ISM frame
$\Gamma_{\rm r}/\Gamma_{\rm s}=3$. 
The parameters are shown in Table 2.
Criterion (II) is satisfied and
the profile classified ``D2''  in Fig. \ref{rhocst}
is seen in this numerical result.
Space-time diagram of shock and rarefaction waves propagations.
RE spreads at the speed $\sim 0.8c$
while FE spreads at the speed $\sim 0.7c$.
}\label{fig:m31}
\end{figure}

\begin{figure}
\plottwo{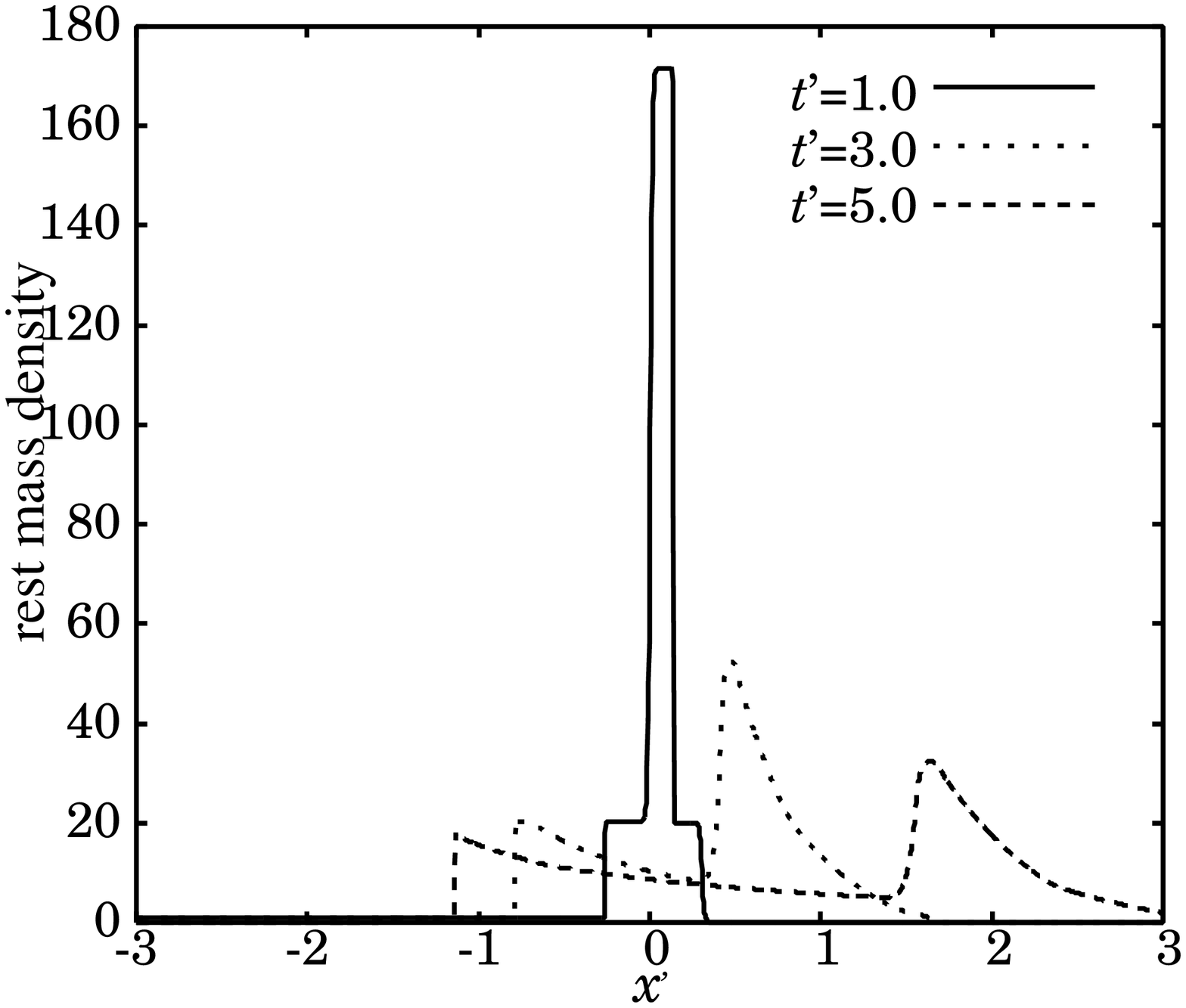}{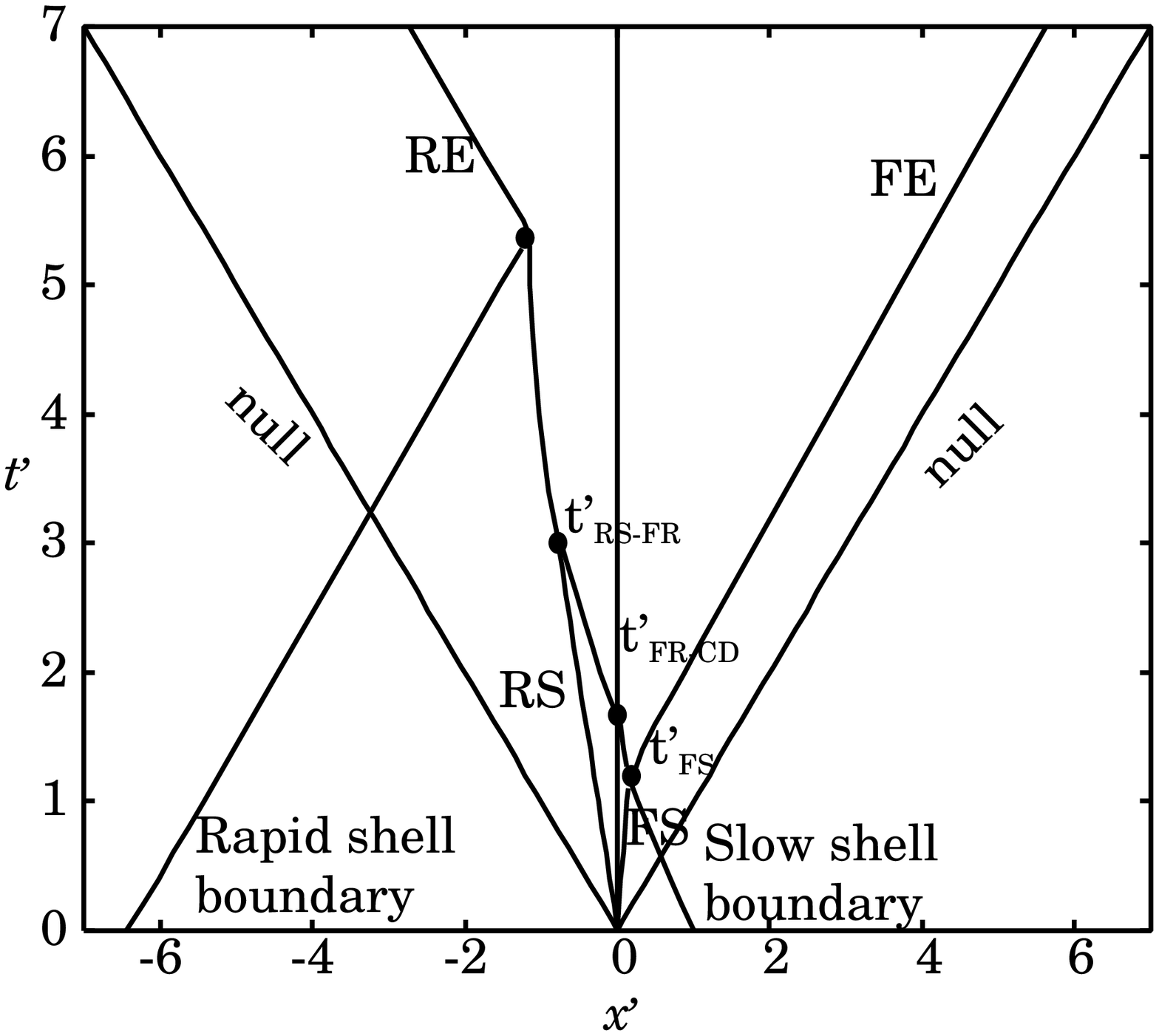}
\caption
%
{
Left:
Time evolution of the 
rest mass density profile in the CD frame for ``equal $m$''. 
In the ISM frame,
$\Gamma_{\rm r}/\Gamma_{\rm s}=20$. 
The parameters are shown in Table 2.
The dilute rapid shell collides with the slow one 
and quickly spreads out. We also see that
the dense slow shock
is pushed forwards by the rapid shell.
Right:
Space-time diagram of shock and rarefaction waves propagations.
RE spreads at the speed $\sim c$
while FE spreads at the speed $\sim 0.9c$.
}\label{fig:m201}
\end{figure}

The time scales for
the ``equal mass'' case are shown
in Fig. \ref{GRB_mass}.
Up to $\Gamma_{\rm r}/\Gamma_{\rm s}\sim 20$,
the criteria (I) and (II) are both satisfied 
and the triple-peaked profile shows up (No. 6 in Table 1).
As $\Gamma_{\rm r}/\Gamma_{\rm s}$ increases,
we obtain
\begin{eqnarray}
\frac{\Delta_{\rm r}^{'}}{\Delta_{\rm s}^{'}}
\sim
\frac{\Gamma_{4}}{\Gamma_{1}},
\qquad
\left|\frac{\beta_{\rm r}^{'}}{\beta_{\rm s}^{'}}\right|
\sim 1  ,
\end{eqnarray}
and
$t_{\rm RS}^{'}$ and $t_{\rm RR-CD}^{'}$
become larger compared with 
$t_{\rm FS}^{'}$ and $t_{\rm FR-CD}^{'}$.
For the  numerical experiment,
we select two cases  
which have 
$\Gamma_{\rm r}/\Gamma_{\rm s}=3$ and $20$.
In each case,
we clearly see the dip corresponding to the criterion (I)
in Figs. 
\ref{fig:m31} and
\ref{fig:m201}.
However, 
as in the ``equal $E$'' case,
the collision of rarefaction waves occurs
in the less dense rapid shell and
the density peak tends to be smoothed out.
Hence we conclude that the equal mass collision with 
a large value of $\Gamma_{\rm r}/\Gamma_{\rm s}$
effectively evolves 
in to the double-peaked (D2) profile. 
In KS01, the authors found a shell-split feature 
in their numerical simulation (Fig. 2 in their paper)
for the  ``equal $m$'' case.
It can be also explained as the D2 profile. 
In Fig. \ref{fig:m201}, 
we see that the rarefaction wave (FR) driven by the breakout
of FS catches up with
the shock wave (RS) from behind at $t^{'}\sim 3$,
since the flow seen from RS is subsonic
in the downstream of RS.
The propagation speed of RS is modified by 
the merge with the rarefaction wave and is
determined by the strengths of the shock wave
and rarefaction wave.
For the current case, 
the propagation speed of RS is almost unchanged 
up to its breakout at $t^{'} \lesssim 5.4$.

The time scales for the 
``equal $\rho$'' case is shown
in Fig. \ref{fig:rho61}. The important point is that
 $\rho_{\rm s}=\rho_{\rm r}$ along CD.
Then the criterion (I) disappears.
In the limit of large
$\Gamma_{\rm r}/\Gamma_{\rm s}$,
we have
\begin{eqnarray}
\frac{\Delta_{\rm r}^{'}}{\Delta_{\rm s}^{'}}
\sim
\frac{1}{3}
\left(\frac{\Gamma_{4}}{\Gamma_{1}}\right)^{2}
\qquad
\left|\frac{\beta_{\rm r}^{'}}{\beta_{\rm s}^{'}}\right|
\sim 
\frac{1}{3}  ,
\end{eqnarray}
and
$t_{\rm RS}$ and $t_{\rm RR-CD}$
become larger compared with 
$t_{\rm FS}$ and $t_{\rm FR-CD}$.
We see this in the numerical experiment
with $\Gamma_{\rm r}/\Gamma_{\rm s}=6$
in Fig. \ref{fig:rho61}.
As in Fig. \ref{fig:m201},
we see also in Fig. \ref{fig:rho61}
that the FR catches up to the RS
and merges
during $3\lesssim t^{'} \lesssim 6.5$.
The last topic is the 
dependence of the above results
on the hitherto fixed model parameters 
$\Gamma_{\rm s}$,
$\Delta_{\rm s}^{'}$, and
$\rho_{\rm s}$.
The results for different  $\Gamma_{\rm s}$ 
are completely the same as in Fig. \ref{GRB}.
The position of each time scale is determined by  
$\Delta_{\rm r}^{'}/\Delta_{\rm s}^{'}$ and
$\beta_{\rm r}^{'}/\beta_{\rm s}^{'}$ 
and they are independent of $\Delta_{\rm s}$  itself.
Hence we omit corresponding figures. 
For different $\rho_{\rm s}$,
the result is also almost unchanged,
since we treat the case of $\Gamma_{\rm r}/\Gamma_{\rm s}>3$ 
for simplicity. Then the sound speed is 
about a few ten percent of the light speed
and has a weak dependence on $\rho_{\rm s}$.
When $\Delta_{\rm s}^{'}$ is increased, 
every time-scale increases linearly 
keeping the relative positions of the time scales.

For a typical blazar,
we set up three slow shell parameters as
$\Gamma_{\rm s}=10$,
$\Delta_{\rm s}^{'}=10^{16}$ cm, and
$\rho_{\rm s}=10^{-26} \rm \ g \ cm^{-3}$.
We show the result in
Fig. \ref{blz} for the ``equal $E$'' case as an example.
The essential difference between GRBs and  
blazars is a typical shell width.
Hence, as explained above,
every time-scale becomes
$10^{6}$ times larger than 
that in Fig. \ref{GRB} 
with the relative positions unchanged.

\subsubsection{Extra case}
Since we have not seen so far
a clear case corresponding to
the criterion (II),
we have performed another specific case to show
what happens for the collision of two 
rarefaction-rarefaction waves collision
(D1 profile in Fig. \ref{rhocst}).
D1 profile is appears most clearly when
the rapid and slow shells have similar
mass densities and 
shell widths in the CD frame.
Hence we do not use the
assumption of $\Delta_{\rm r}=\Delta_{\rm s}$
here only.
Instead, we employ the condition that
$\rho_{\rm r}=\rho_{\rm s}$,
$\Delta^{'}_{\rm r}=\Delta^{'}_{\rm s}$, and
$\Gamma_{\rm r}/\Gamma_{\rm s}=20$. 
In Fig. \ref{fig:RRFR},  
D1 profile is indeed produced.
A larger value of
$\Gamma_{\rm r}/\Gamma_{\rm s}$ 
produce a greater dip than that 
shown in Fig. \ref{fig:RRFR}.

\subsection{Shell Spreading}

In principle,
we can obtain
the speed of rarefaction wave using Riemann invariants 
(e.g., Zel'dovich \& Raizer, 1966).
In the relativistic limit,
it is known that the speed of the head  
of rarefaction wave is close to the speed of light (e.g., Anile 1989).
As the EOS of the shocked region deviates 
from the relativistic one, the speed is reduced from the light speed
and the intermediate regime
should be treated by numerical calculations  
(e.g., Wen, Panaitescu, \& Laguna 1997).
It is worthwhile to note that from the values 
$\Gamma_{\rm r}^{'}$ and
$\Gamma_{\rm s}^{'}$ in Table 2 
we see that the EOS in the forward-shocked region is 
a non-relativistic one
while the  reverse-shocked region 
extends from the non-relativistic to the
relativistic regime
for the parameter ranges we adopted.

Numerical results for  
shell spreading are shown 
in Figs. 
\ref{fig:e31}, 
\ref{fig:e61}, 
\ref{fig:m31}, 
\ref{fig:m201}, and
\ref{fig:rho61},
in which the width of the shell may be described as
\begin{eqnarray}
\Delta_{\rm tot}^{'}(t^{'})
\simeq
\Delta_{\rm s,FS}^{'}+\Delta_{\rm r,RS}^{'}
+v_{\rm FE}(t^{'}-t_{\rm FS}^{'})+v_{\rm RE} (t^{'}-t_{\rm RS}^{'}) ,
\end{eqnarray}
where 
$\Delta_{\rm tot}^{'}$ is the total width of the shell
in the CD frame.
We should stress that although
a lot of authors assume that 
the shell width is not changed after collisions
(e.g., Spada et al. 2000; NP02) for simplicity,
the reality is that the shells spread
at $v_{\rm FE}$ and  $v_{\rm RE}$ in the
forward and backward directions, respectively.

\subsection{Energy Conversion Efficiency}\label{sec:eff}

The conversion efficiency
of the bulk kinetic energy to the internal one
is one of the most 
important issues to explore the nature of 
the central engine of relativistic outflows 
and many authors have studied it
(e.g., Kumar 1999; Tanihata et al. 2002).

\subsubsection{Two-mass-collision model}

Let us briefly review the widely-used
two-mass-collision model
(e.g., Piran 1999).
From the momentum and energy conservations, we have
\begin{eqnarray}\label{eq:hyd}
m_{\rm r}\Gamma_{\rm r}+m_{\rm s}\Gamma_{\rm s}
&=&
(m_{\rm m}+{\cal E}_{\rm m}/c^{2}) \Gamma_{\rm m},\nonumber \\
m_{\rm r}\Gamma_{\rm r}\beta_{\rm r}+
m_{\rm s}\Gamma_{\rm s}\beta_{\rm s}
&=&
(m_{\rm m}+{\cal E}_{\rm m}/c^{2}) 
\Gamma_{\rm m}\beta_{\rm m},
\end{eqnarray}
where 
$m_{\rm m}=m_{\rm r}+m_{\rm s}$,
${\cal E}_{\rm m}={\cal E}_{\rm r}+{\cal E}_{\rm s}$, and 
$\Gamma_{\rm m}$ 
are 
the mass,
the internal energy, and
the Lorentz factor of
the merged shell, respectively, and
$m_{\rm r}$ 
and 
$m_{\rm s}$ 
are the rest mass of the rapid and slow shells,
respectively. 
Then we obtain the efficiency 
$\epsilon$ as
\begin{eqnarray}
\epsilon=
1-\frac{(m_{\rm r}+m_{\rm s})\Gamma_{\rm m}}
{m_{\rm r}\Gamma_{\rm r}
+m_{\rm s}\Gamma_{\rm s}},
\qquad
\Gamma_{\rm m}^{2}=
\Gamma_{\rm r}
\Gamma_{\rm s}
\frac{m_{\rm r}\Gamma_{\rm r}+m_{\rm s}\Gamma_{\rm s}}
{m_{\rm r}\Gamma_{\rm s}
+m_{\rm s}\Gamma_{\rm r}}  .
\end{eqnarray}
%
It is a useful shortcut to approximate
$\Gamma_{\rm m}\sim\Gamma_{2}=\Gamma_{3}$
without solving Eq. (\ref{fblazar}).
Using this shortcut,
for the ``equal mass density'' case 
($m_{\rm r}/\Gamma_{\rm r}=m_{\rm s}/\Gamma_{\rm s}$),
we have $\Gamma_{\rm m}^{2}=(\Gamma_{\rm r}^{2}+\Gamma_{\rm s}^{2})/2$.
For the ``equal mass'' case ($m_{\rm r}=m_{\rm s}$),
we have $\Gamma_{\rm m}^{2}=\Gamma_{\rm r}\Gamma_{\rm s}$.
For ``equal energy'' 
($m_{\rm r}\Gamma_{\rm r}=m_{\rm s}\Gamma_{\rm s}$),
we have 
$\Gamma_{\rm m}^{2}= 
2(\Gamma_{\rm s}^{2}\Gamma_{\rm r}^{2})
/(\Gamma_{\rm s}^{2}+\Gamma_{\rm r}^{2})$. 
Then, the efficiency $\epsilon$
in each case is given by
\begin{eqnarray} \label{eq:eff-ana}
\epsilon
&\simeq&
\left\{ 
\begin{array}{ll}
1-
\frac{1}{\sqrt{2}}
\left(1+
\frac{\Gamma_{\rm s}}{\Gamma_{\rm r}}\right)
\left[1+
\left(
\frac{\Gamma_{\rm s}}{\Gamma_{\rm r}}\right)^{2}
\right]^{-1/2}
& 
({\rm equal} \ \rho)  ,\\
1- \left(\frac{4}{2+\frac{\Gamma_{\rm r}}{\Gamma_{\rm s}}
+\frac{\Gamma_{\rm s}}{\Gamma_{\rm r}}}\right)^{1/2} 
&  
({\rm equal} \ m) ,\\
1-
\frac{1}{\sqrt{2}}
\left(1+
\frac{\Gamma_{\rm r}}{\Gamma_{\rm s}}\right)
\left[1+
\left(
\frac{\Gamma_{\rm r}}{\Gamma_{\rm s}}\right)^{2}
\right]^{-1/2}
& 

({\rm equal} \ E) . \\
\end{array}
\right.
\end{eqnarray}
As the value of $\Gamma_{r}/\Gamma_{s}$
gets larger for the ``equal $m$'' case, 
the efficiency becomes larger and
approaches $\sim 1$.
For ``equal $E$'' and ``equal $\rho$'' cases,
it goes asymptotically to $\sim 0.3$.
Thus we find that ``equal $m$'' is the most effective
collision if we neglect the rarefaction waves. 

\subsubsection{Shock model}

The consideration of
shock dynamics provides us with the information of
the assignment of the total dissipation energy
$E_{\rm diss}$ to the 
forward- and reverse-shocked regions 
$E_{\rm FS}$ and $E_{\rm RS}$.
From the equations of mass continuity in the CD frame, we have
$
\frac{\Delta_{\rm s,FS}^{'}}
{\Delta_{\rm s}^{'}}=
\frac{\Gamma_{1}^{'}(\hat{\gamma}_{2}-1)}
{\hat{\gamma}_{2}\Gamma_{1}^{'}+1}$ and
$
\frac{\Delta_{\rm s,RS}^{'}}
{\Delta_{\rm r}^{'}}=
\frac{\Gamma_{4}^{'}
(\hat{\gamma}_{3}-1)}
{\hat{\gamma}_{3}\Gamma_{4}^{'}+1}
$   
Assuming a large value of $\Gamma_{4}/\Gamma_{1}$,
then we have $e_{2}+P_{2}\simeq e_{3}+P_{3}$ and
the dissipated energies are mainly controlled by shell widths.
They are written as
\begin{eqnarray} 
\frac{E_{\rm FS}}{E_{\rm diss}}
\simeq
\frac{\Delta_{\rm s,FS}^{'}}
{\Delta_{\rm s,FS}^{'}+\Delta_{\rm r,RS}^{'}}
,\quad
\frac{E_{\rm RS}}{E_{\rm diss}}
\simeq
\frac{\Delta_{\rm r,RS}^{'}}
{\Delta_{\rm s,FS}^{'}+\Delta_{\rm r,RS}^{'}}
\end{eqnarray}   
where 
$E_{\rm FS}$ and $E_{\rm RS}$, 
internal energy of forward and reverse shocked regions
just after the shock crosses each shell, respectively
(e.g., Spada et al. 2000).
However, 
the values
$E_{\rm FS}/E_{\rm diss}$ and
$E_{\rm RS}/E_{\rm diss}$
will begin to deviate
from the the above approximation 
when the rarefaction waves start to propagate,
since they reconvert the 
internal energy into the kinetic one.
The result including the
rarefaction waves is shown below.

\subsubsection{Numerical Results}

The estimation
of the energy conversion
efficiencies with shock and rarefaction waves
taken into account are presented in 
Figs. \ref{fig:effene}, \ref{fig:effmass}
and \ref{fig:effrho}
based on the numerical simulations.
From the analogy of the two-mass-collision model,
we define the efficiency measured in the ISM frame as
\begin{eqnarray}\label{eq:eff}
\epsilon(t)&\equiv&1-
\frac{ \int 
\Gamma(t,x)~dm(t,x)}
{\Gamma_{\rm r}m_{\rm r}
+\Gamma_{\rm s}m_{\rm s}
}\nonumber \\
&=&
1-\frac{ \int 
\Gamma(t,x)~\rho(t,x)
\Gamma(t,x) dx}
{\rho_{\rm r}\Delta_{r}\Gamma_{\rm r}^{2}
+\rho_{\rm s}\Delta_{s}\Gamma_{\rm s}^{2}
}  ,
\end{eqnarray}
where
$dm(t,x)$,
$\rho(t,x)$,
$\Gamma(t,x)$, and
$\Gamma(t,x) dx$,
are
the rest mass element,
the rest mass density, and
the Lorentz factor, 
seen in the ISM frame, 
and 
the length of line element in the CD frame,
respectively.
We assume the origins of both frames coinside
with each other at $t=t^{'}=0$.
The Lorentz transformation of 
$\Gamma(t,x)$ is wtitten as
$\Gamma(t,x)
=\Gamma_{\rm CD}\Gamma^{'}(t,x)
[1-\beta_{\rm CD}\beta^{'}(t,x)]$.
The hypersurface of a constant time in the ISM frame
does not coinside with that in CD frame.
Hence, to evaluate spatial integrations 
at a certain time in the ISM frame,
we must collect the values of physical quantities in the 
inegrand for different times in the CD frame.
Unfortunately it is technically difficult task to 
perform this. Hence we report to an approximation that
$\rho(t,x)$ and
$\Gamma(t,x)$ are replaced by 
$\rho(t^{'},x^{'})$ and
$\Gamma(t^{'},x^{'})$, respectively.
This is only valid near the original point 
of the CD frame and, 
elsewhere, mixes up those quantities at
the different time slice.
We believe, however, that this still gives the behavior of
the efficiency and 
the essencial role of rarefaction waves.
In these figures, we compare the numerical results 
with the prediction by the
two-mass-collision approximation.  
We find that the two-mass-collision model
well reproduces the hydrodynamical
results just before 
rarefaction waves begin to propagate.
After the shock waves break-out the shells,
the conversion effeciency
is reduced by several ten percent from 
the estimate of the two-mass-collision model 
after several dynamical times.
It is noted again 
that no cooling effect is taken into account here.

\section{SUMMARY AND DISCUSSION}

In this paper
we have illuminated 
the difference between the simple two-mass-collision model
and the hydrodynamical one of the internal shock.
We have studied 
1D hydrodynamical simulations of the two-shell-collisions  
in the CD frame 
taking the shock and rarefaction waves into account.
Below we summarize our results and give some discussions.

(1)
By comparing the relevant time scales of shock and rarefaction 
waves,
we have completely classified 
the evolutions
of the two-shell-collisions using
six physical parameters, that is,
the widths, 
rest mass densities, 
and velocities
of the two colliding shells.
We find that rarefaction waves
have a significant effect on the dynamics.
In principle, the rest mass density 
profile can be evolved into single-, double-, 
and triple-peaked features.
In the limit of 
$\Gamma_{\rm r}, \Gamma_{\rm s}\gg1$,
the features are essentially  
characterized by only three
parameters: the ratios of 
Lorentz factors,
widths, and 
rest mass densities.
The combination of the values of 
$\Gamma_{\rm r}/\Gamma_{\rm s}$
and
$\rho_{\rm r}/\rho_{\rm s}$ 
determines
the relative orders of 
the time scales of various wave propagations,
while the value of
$\Delta_{\rm r}/\Delta_{\rm s}$ 
controls normarizations of the time scales.

(2)
Bearing in mind the application
to relativistic outflows such as
GRBs and blazars,
we specifically examine the cases of
``equal $\rho$'',
``equal $m$'',
and ``equal $E$''.
For the ``equal $\rho$'' case, 
the profile is single-peaked.
The rarefaction wave produced when 
the FS breaks out reaches
CD and then catches up with RS.
In the case of ``equal $m$'', the profile should 
in principle become
triple-peaked according to our classification scheme.
In practice, however, there is 
very little time for the FR-RR
collision to make a clear dip, 
while there is 
a lot more time for the FR
to create a dip for a fairly wide range of parameters. 
Therefore, the profile in this case
is effectively double-peaked.
For the ``equal $E$'' case, 
the profile is classified as triple-peak.
However, again, there is 
very little time for the FR-RR
collision to make a dip.
A very large mass-density gradient between
forward- and reverse-shocked 
regions  makes the dip even less pronounced.
Furthermore,
there is again little time for FR
to create a dip for a 
fairly wide range of parameters. 
Hence, we conclude that  the profile for the  ``equal $E$''
is effectively single-peakd.
If the cooling time scale 
is sufficiently long in the shocked region, 
electromagnetic radiations 
will show these profiles.

(3)
For large $\Gamma_{\rm r}/\Gamma_{\rm s}$,
we have shown that the spreading velocity of the shells
after collision is close to the speed of light.
Hence, the often used approximation
of constant shell width after collision
is not very good in treating multiple collisions.
For examle, 
the authors in NP02 claim that
the ``equal energy'' case is suggested
for the shell Lorentz factors in GRBs,
assuming that $L> \Delta$, 
where $L$ is the separation 
distance between two shells.
If the interval of 
the first and the second collisions is long,
however,
the shell spreading effect
cannot be ignored and
the case of $L< \Delta$ 
should be included  
in the analysis.
Then the difference between 
the ``equal $m$'' and ``equal $E$'' cases
might be wiped away.

(4)
As the shell spreads after collision, the internal energy 
is converted back to the bulk kinetic energy 
due to thermal expansion.
We have numerically studied the time-dependent 
energy conversion efficiency quantitatively.
Since we have neglected cooling processes, 
the conversion effciency rises up to the order of unity.
This should be corresponding to the event 
in the regime of ``weak cooling'' (Kino \& Takahara 2004).
If $\Gamma_{\rm r}/\Gamma_{\rm s}\gg1$
and the time-interval between collisions is long,
the conversion efficiency  
will be substantially deviated the estimate of
the two-mass-collision model.

We appreciate the insightful comments 
and suggestions of the referee.
M.K. thank S. Kobayashi, K. Asano 
and F. Takahara for useful remarks and discussions.  
A.M. acknowledges support from Japan Sosiety for the
Promotion of Science (JSPS).
This work is also supported in part by 
the Grant-in-Aid Program for Scientific Research
(14340066, 14740166 and 14079202) from the Ministry 
of Education, Science, Sports, and Culture of Japan.


\clearpage


\begin{deluxetable}{rcrr}
\tablecaption{Various types of the evolution}
\tablehead{
\colhead{No.}
& \colhead{$\rho$} 
& \colhead{timescale$^{a}$}
& \colhead{profile$^{b}$}}
\startdata
1 
& $\rho_{\rm s}>\rho_{\rm r}$
& $t_{\rm RS}^{'}<t_{\rm RR-CD}^{'}< t_{\rm RR-FS}^{'}$ 
& S
\\
2
&
& $t_{\rm RS}^{'}<t_{\rm RR-CD}^{'}< t_{\rm FS}^{'}< t_{\rm RR-FR}^{'}$ 
& D1 
\\
3
&
& $t_{\rm RS}^{'}<t_{\rm FS}^{'}< t_{\rm RR-CD}^{'}< t_{\rm RR-FR}^{'}$ 
& D1 
\\
4
& 
& $t_{\rm RS}^{'}<t_{\rm FS}^{'}< t_{\rm FR-CD}^{'}< t_{\rm RR-FR}^{'}$ 
& T 
\\
5
&
& $t_{\rm FS}^{'}<t_{\rm FR-CD}^{'}< t_{\rm FR-RS}^{'}$ 
& D2
\\
6
&
& $t_{\rm FS}^{'}<t_{\rm FR-CD}^{'}< t_{\rm RS}^{'}< t_{\rm RR-FR}^{'}$ 
& T 
\\
7
&
& $t_{\rm FS}^{'}<t_{\rm RS}^{'}<_{\rm FR-CD}^{'}< t_{\rm RR-FR}^{'}$ 
& T 
\\
8
&
& $t_{\rm FS}^{'}<t_{\rm RS}^{'}<_{\rm RR-CD}^{'}< t_{\rm RR-FR}^{'}$ 
& D1 
\\
\hline
9
&
$\rho_{\rm s}<\rho_{\rm r}$
&$t_{\rm RS}^{'}<t_{\rm RR-CD}^{'}< t_{\rm RR-FS}^{'}$ 
& D2
\\
10
&
&$t_{\rm RS}^{'}<t_{\rm RR-CD}^{'}< t_{\rm FS}^{'}< t_{\rm RR-FR}^{'}$ 
& T 
\\
11
&
&$t_{\rm RS}^{'}<t_{\rm FS}^{'}< t_{\rm RR-CD}^{'}< t_{\rm RR-FR}^{'}$ 
& T 
\\
12
&
&$t_{\rm RS}^{'}<t_{\rm FS}^{'}< t_{\rm FR-CD}^{'}< t_{\rm RR-FR}^{'}$ 
& D1 
\\
13
&
& $t_{\rm FS}^{'}<t_{\rm FR-CD}^{'}< t_{\rm FR-RS}^{'}$ 
& S
\\
14
&
& $t_{\rm FS}^{'}<t_{\rm FR-CD}^{'}< t_{\rm RS}^{'}< t_{\rm RR-FR}^{'}$ 
& D1 
\\
15
&
& $t_{\rm FS}^{'}<t_{\rm RS}^{'}<t_{\rm FR-CD}^{'}
<t_{\rm RR-FR}^{'}$ 
& D1
\\
16
&
& $t_{\rm FS}^{'}<t_{\rm RS}^{'}<t_{\rm RR-CD}^{'}
<t_{\rm RR-FR}^{'}$ 
& T 
\\
\hline
17
&
$\rho_{\rm s}=\rho_{\rm r}$
&$t_{\rm RS}^{'}< t_{\rm RR-FS}^{'}$ 
& S
\\
18
&
&$t_{\rm RS}^{'}<t_{\rm FS}^{'}< t_{\rm RR-FR}^{'}$ 
& D1 
\\
19
&
& $t_{\rm FS}^{'}< t_{\rm FR-RS}^{'}$ 
& S
\\
20
&
& $t_{\rm FS}^{'}<t_{\rm RS}^{'}< t_{\rm RR-FR}^{'}$ 
&  D1
\enddata

\tablenotetext{a}{Notations are given in \S 2.2.}
\tablenotetext{b}{Representative profiles of
S, D1,  D2, and T are schematically 
shown in Fig. 2. }

\end{deluxetable}

\begin{deluxetable}{lcrlrrr}
\tablecaption{Parameter sets for numerical simulations 
of ``equal $E$'', ``equal $m$'' and
``equal $\rho$''.}
\tablehead{
\colhead{No.$^{a}$}
& \colhead{$\Gamma_{\rm r}/\Gamma_{s}$ }
& \colhead{$\Gamma_{\rm CD}$$^{b}$}
& \colhead{$\Gamma_{\rm r}^{'}$}
& \colhead{$\Gamma_{\rm s}^{'}$}
& \colhead{$\Delta_{\rm r}^{'}/\Delta_{\rm s}^{'}$}
& \colhead{$\rho_{\rm s}/\rho_{\rm r}$}
}
\startdata
4(equal $E$)
& 3 
& 6.7
& 1.34
& 1.05
& 2.3
& 9
\\
4(equal $E$)$^{c}$
& 6 
& 7.0
& 2.26
& 1.06
& 2.8
& 36
\\
6(equal $m$)
& 3 
& 7.6
& 1.25 
& 1.09
& 2.6
& 3
\\
5(equal $m$)$^{c}$
& 20 
& 11.8
& 4.29
& 1.40
& 6.4
& 20
\\
19(equal $\rho$)
& 6 
& 12.2
& 1.43
& 1.43
& 6.0
& 1
\\
--(equal $\rho$)$^{c}$ $^{d}$
& 20 
& 22.4
& 2.35
& 2.35
& 1.0
& 1
\enddata

\tablenotetext{a}{These numbers correspond 
to those shown in Table 1.}

\tablenotetext{b}{The values are obtained 
by solving Eq. (\ref{fblazar}).
The slow shell Lorentz factor is fixed as $\Gamma_{\rm s}=5$. }

\tablenotetext{c}{ We use 
$\hat{\gamma}_{3}=4/3$.}

\tablenotetext{d}{ This corresponds to the 
case D1 profile in Fig. 2 without the assumption of 
$\Delta_{\rm s}=\Delta_{\rm r}$.}

\end{deluxetable}


\clearpage

\begin{figure}
\plotone{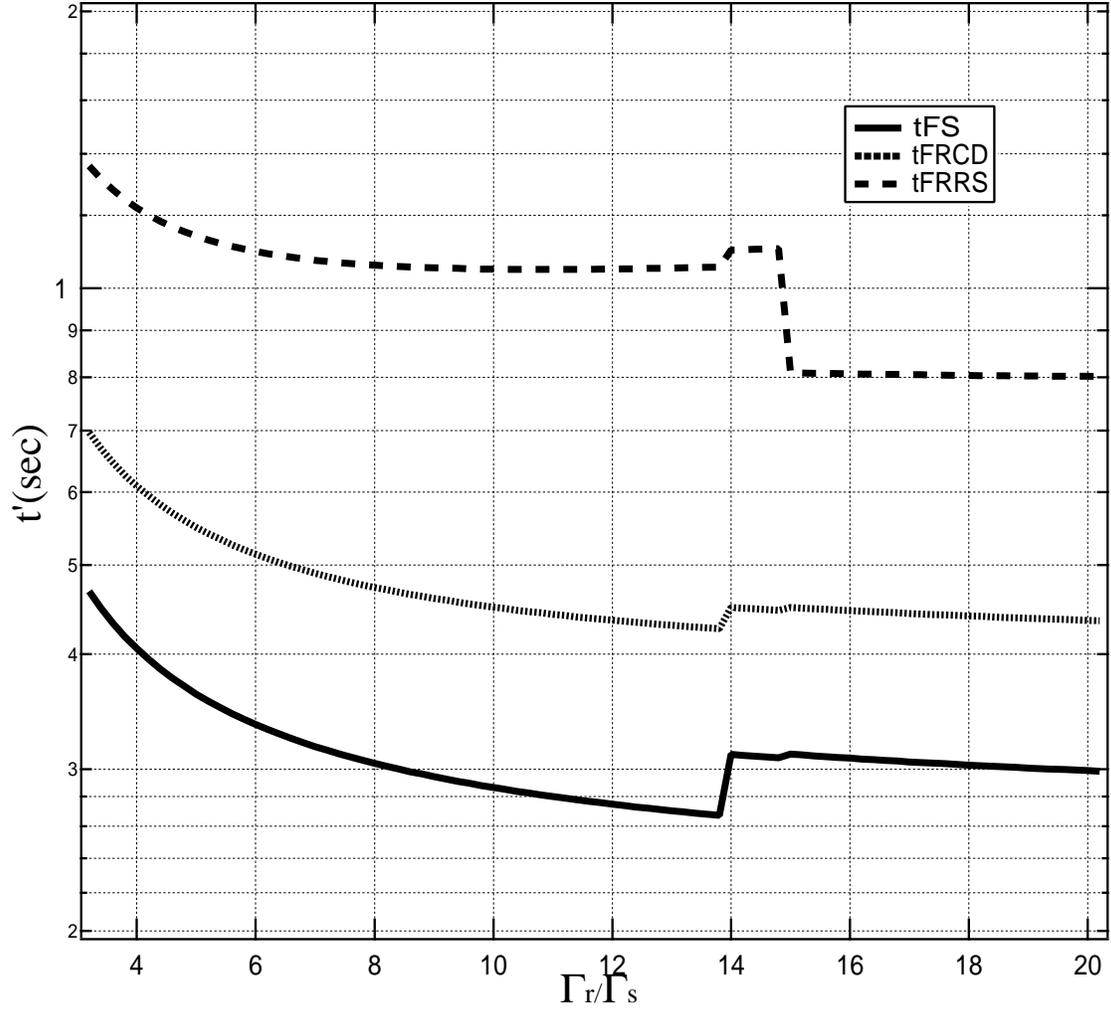}
\caption
%
{Time scales for ``equal $\rho$''. 
Case 19 in Table 1 is realized in the whole range.
The slight jumps of the time scales 
at $\Gamma_{\rm r}/\Gamma_{\rm s}\sim 14, 15$
correspond to the abrupt change of adiabatic index.
The softening of EOS in the relativistic
regime gives slower shock waves
and the time scales get longer accordingly.}
\label{GRBrho}
\end{figure}

\begin{figure}
\plottwo{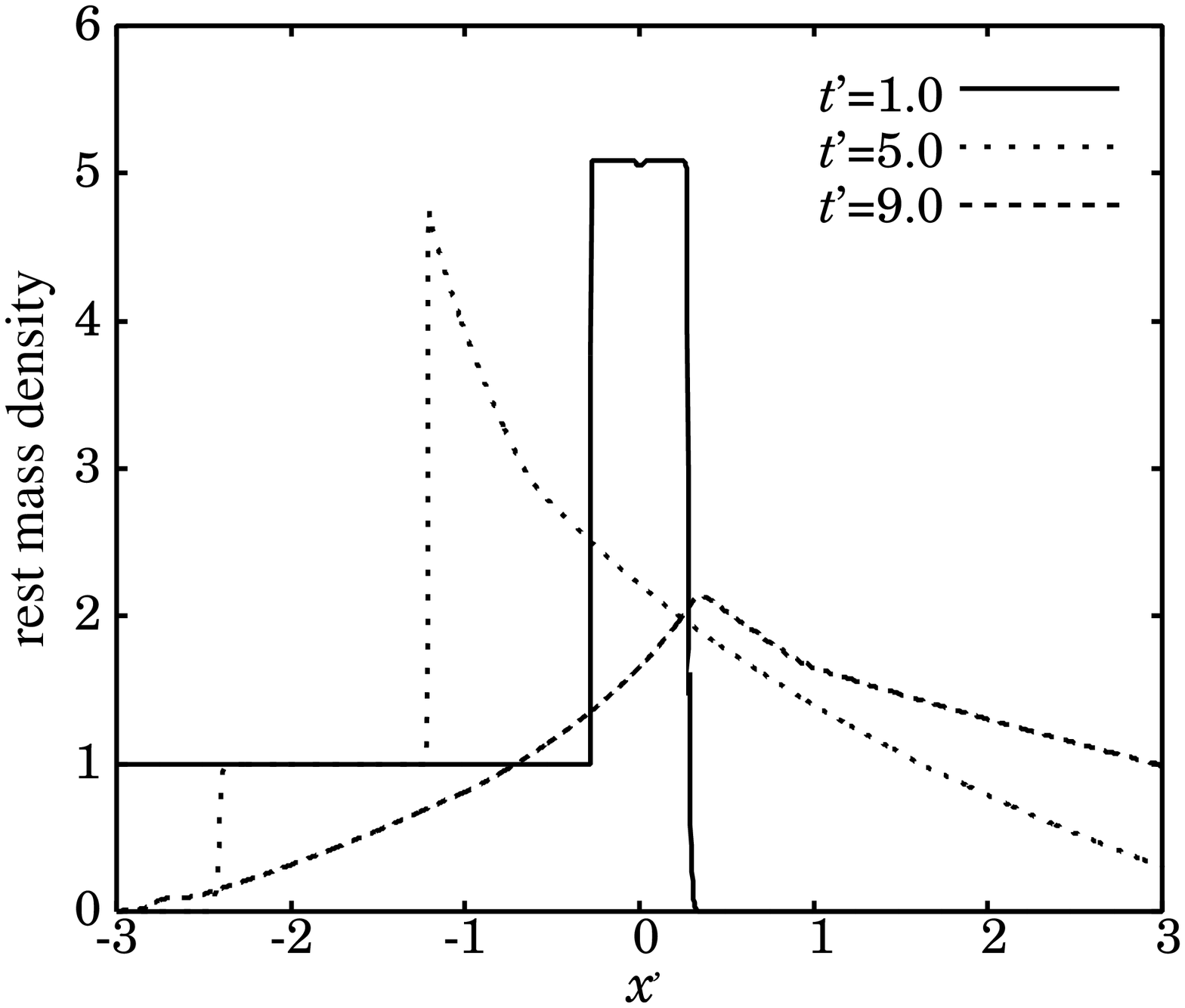}{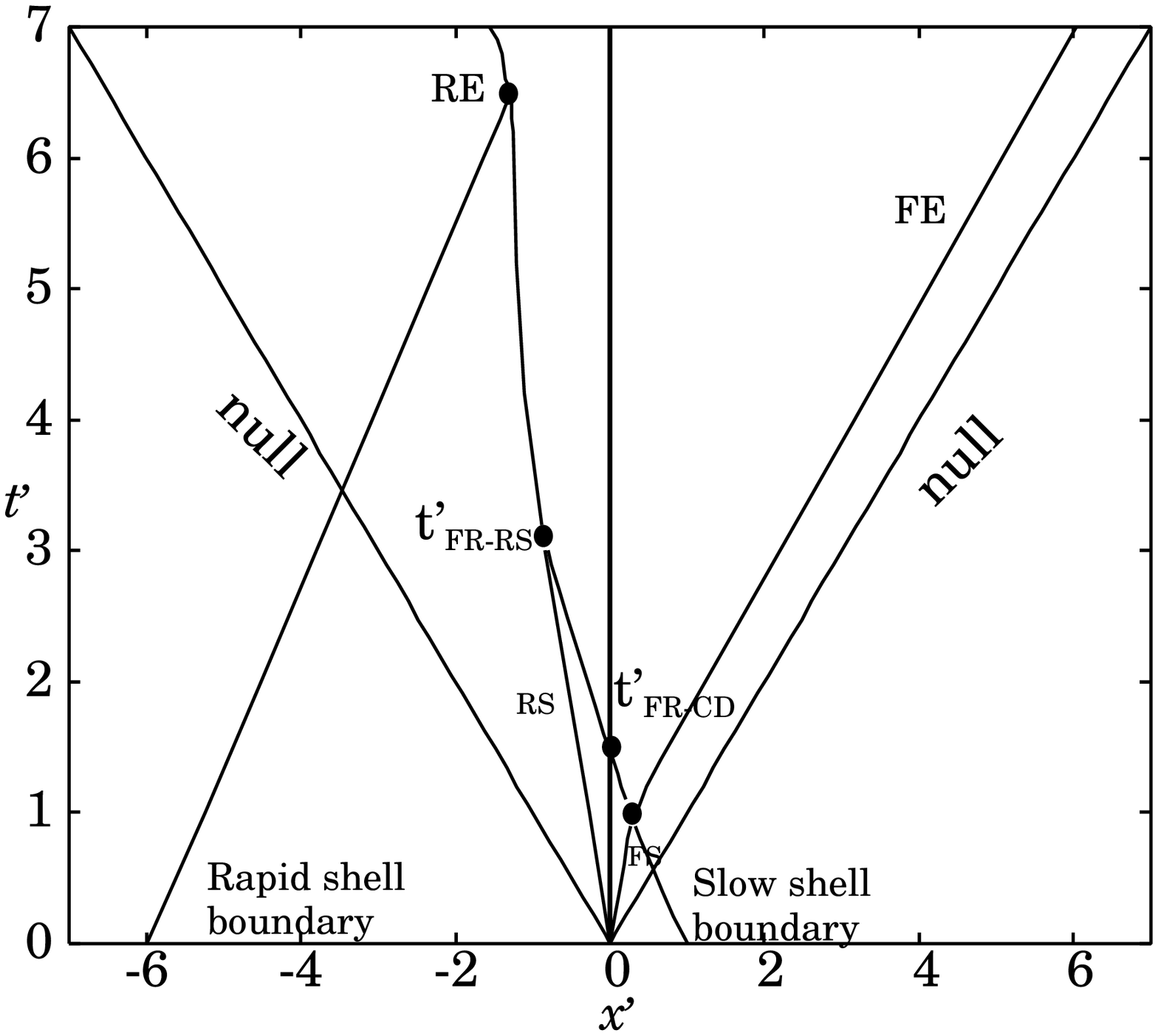}
\caption
%
{
Left:
Time evolution of the 
rest mass density profile in the CD frame for ``equal $\rho$''. 
In the ISM frame,
$\Gamma_{\rm r}/\Gamma_{\rm s}=6$. 
The parameters are shown in Table 2.
Right: 
Space-time diagram of shock and rarefaction waves propagations.
RE spreads at the speed $\sim c$
while FE spreads at the speed $\sim 0.8c$.
}\label{fig:rho61}
\end{figure}

\begin{figure}
\plotone{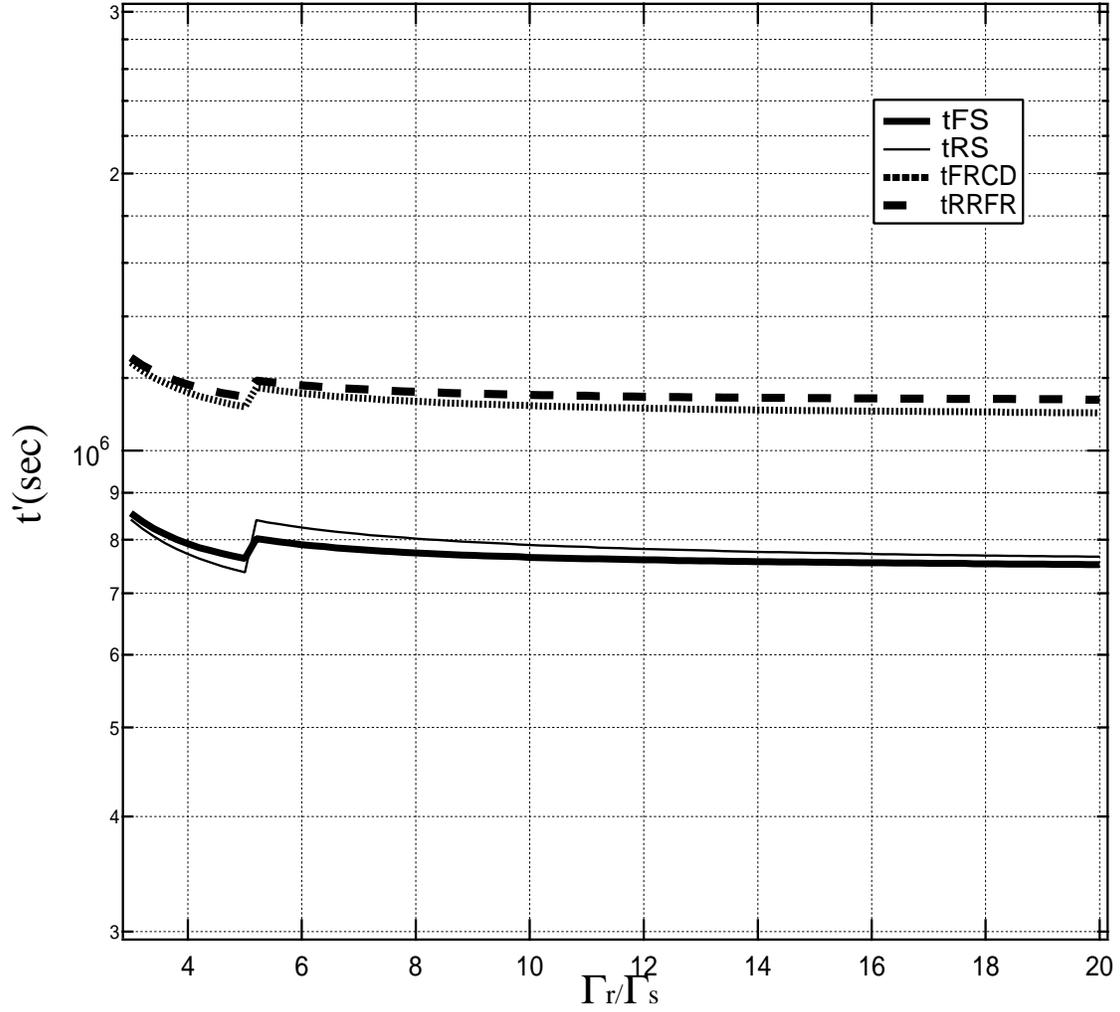}
\caption
%
{The same as Fig. \ref{GRB} but for blazars.
Compared with Fig. \ref{GRB} 
the relative positions of various time scales are 
almost unchanged
while the absolute value of each time-scale is reduced by
a factor of $10^{6}$ which is the ratio of the shell
width of  GRB to that of blazar.
}\label{blz}
\end{figure}

\begin{figure}
\plottwo{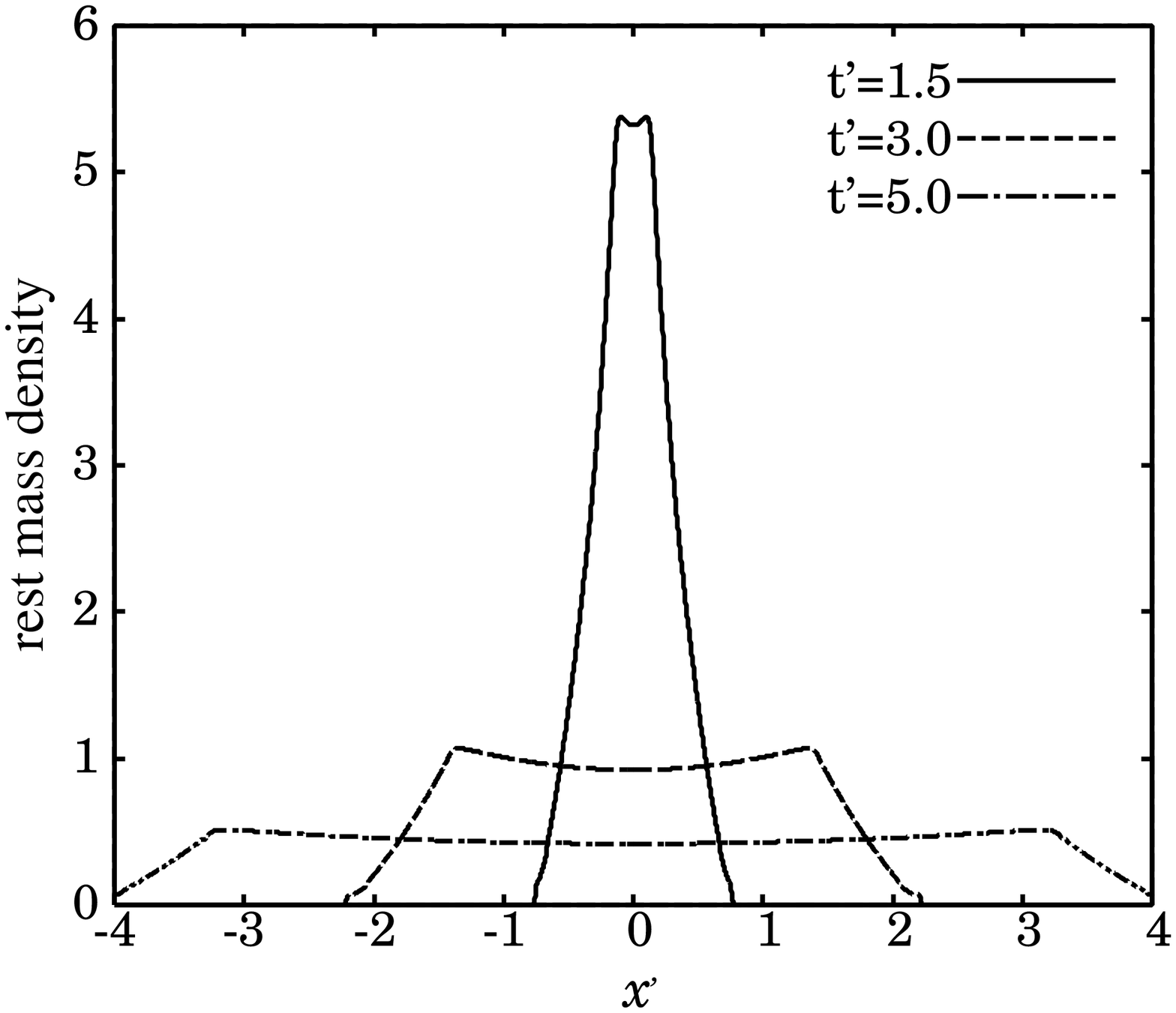}{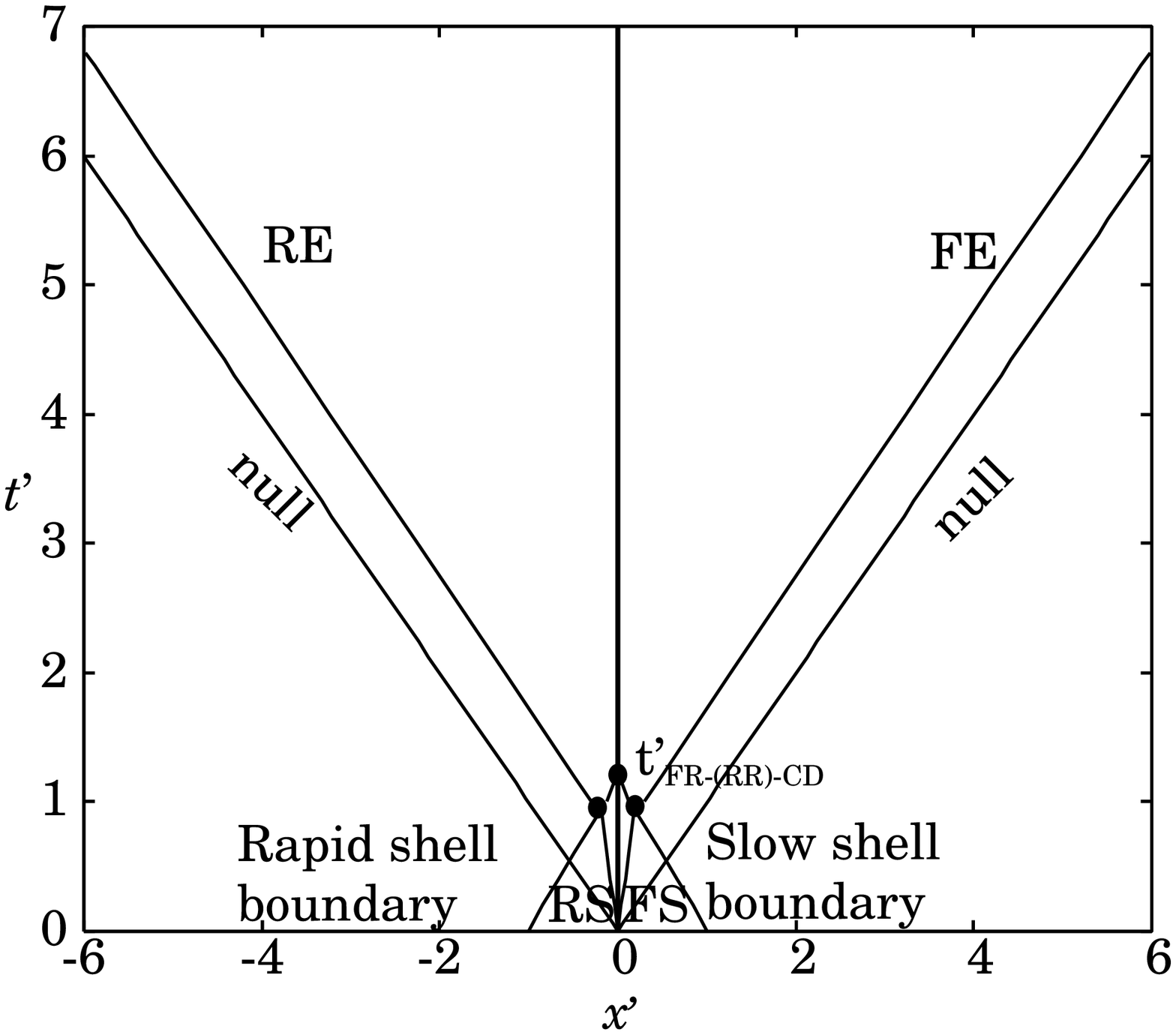}
\caption
%
{Left:
Time evolution of the 
rest mass density profile in the CD frame for ``equal $\rho$''. 
In the ISM frame,
$\Gamma_{\rm r}/\Gamma_{\rm s}=20$. 
The parameters are shown in Table 2.
By the rarefaction-rarefaction wave collision,
the bump is generated at the center of the shell.
This corresponds to the D1 profile in Fig. \ref{rhocst}.
Right:
Space-time diagram of shock and rarefaction waves propagations.
Both RE and FE spread at the speed $\sim c$.
}\label{fig:RRFR}
\end{figure}

\begin{figure}
\plotone{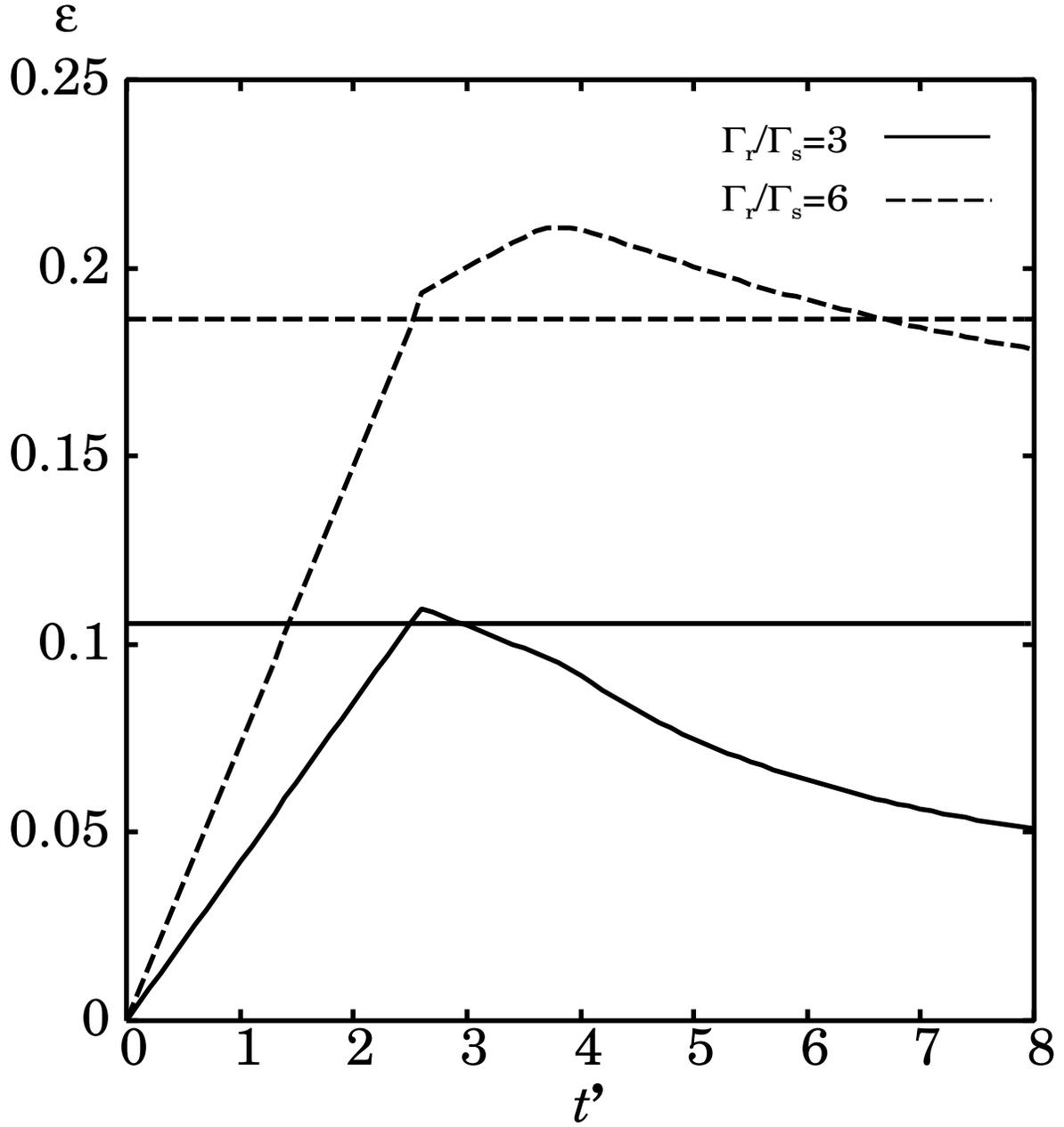}
\caption
%
{Left:
Time evolutions of 
the conversion efficiency defined by Eq. (\ref{eq:eff})
for the ``equal energy'' case.
During shock propagations, 
$\epsilon$ 
approaches the two-mass-collision estimate given by
Eq. (\ref{eq:eff-ana}) for each case.
As the rarefaction waves begin to propagate,
the efficiencies are reduced by them.
This suppression  is can be ascribed 
to the thermal expansion.}\label{fig:effene}
\end{figure}
\begin{figure}
\plotone{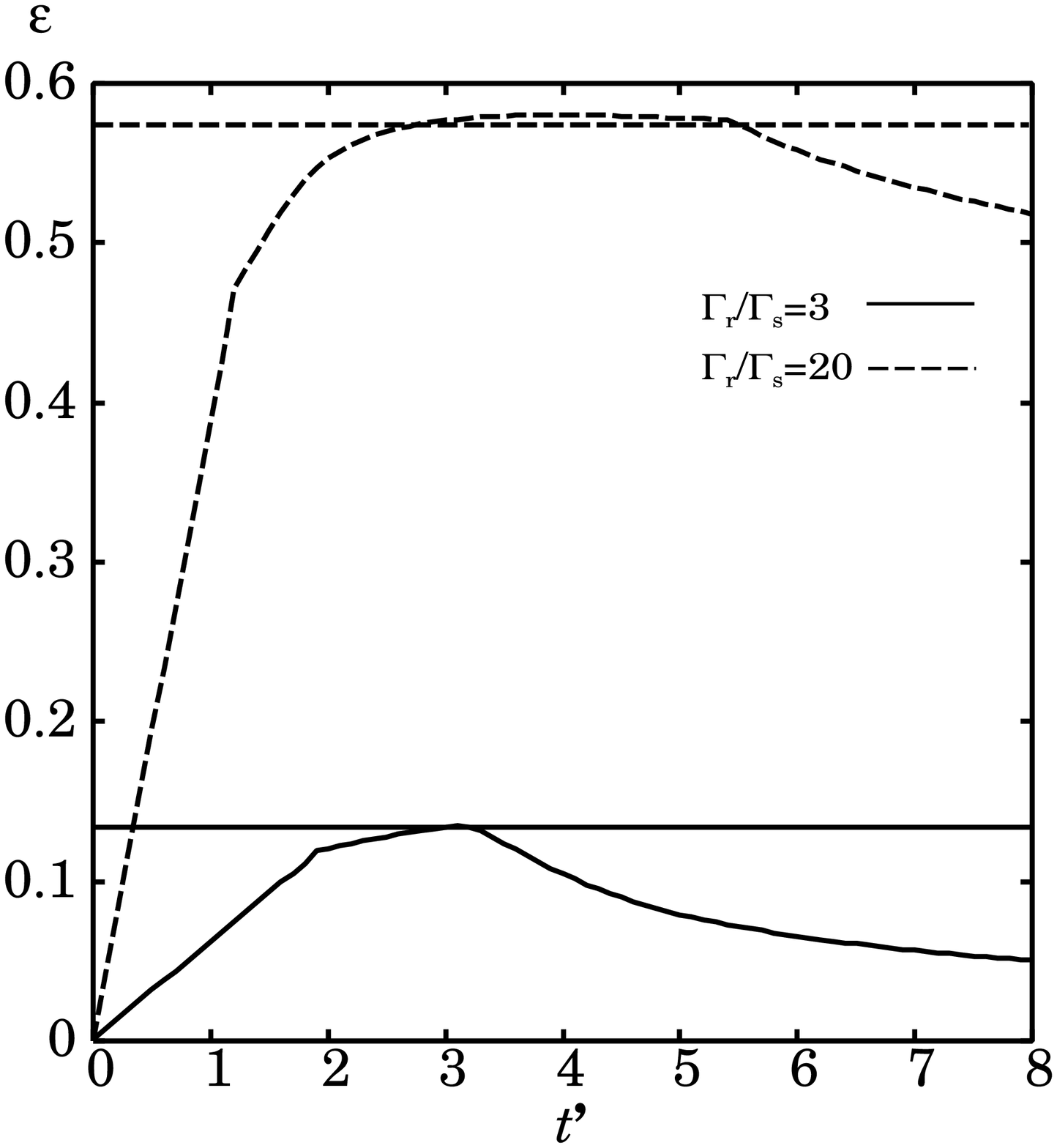}
\caption
%
{Time evolutions of the  
conversion efficiency defined by Eq. (\ref{eq:eff})
for the ``equal $m$'' case.
During shock propagations, 
$\epsilon$ 
approaches the two-point-mass estimate given by
Eq. (\ref{eq:eff-ana}) for each case.
}\label{fig:effmass}
\end{figure}
\begin{figure}
\plotone{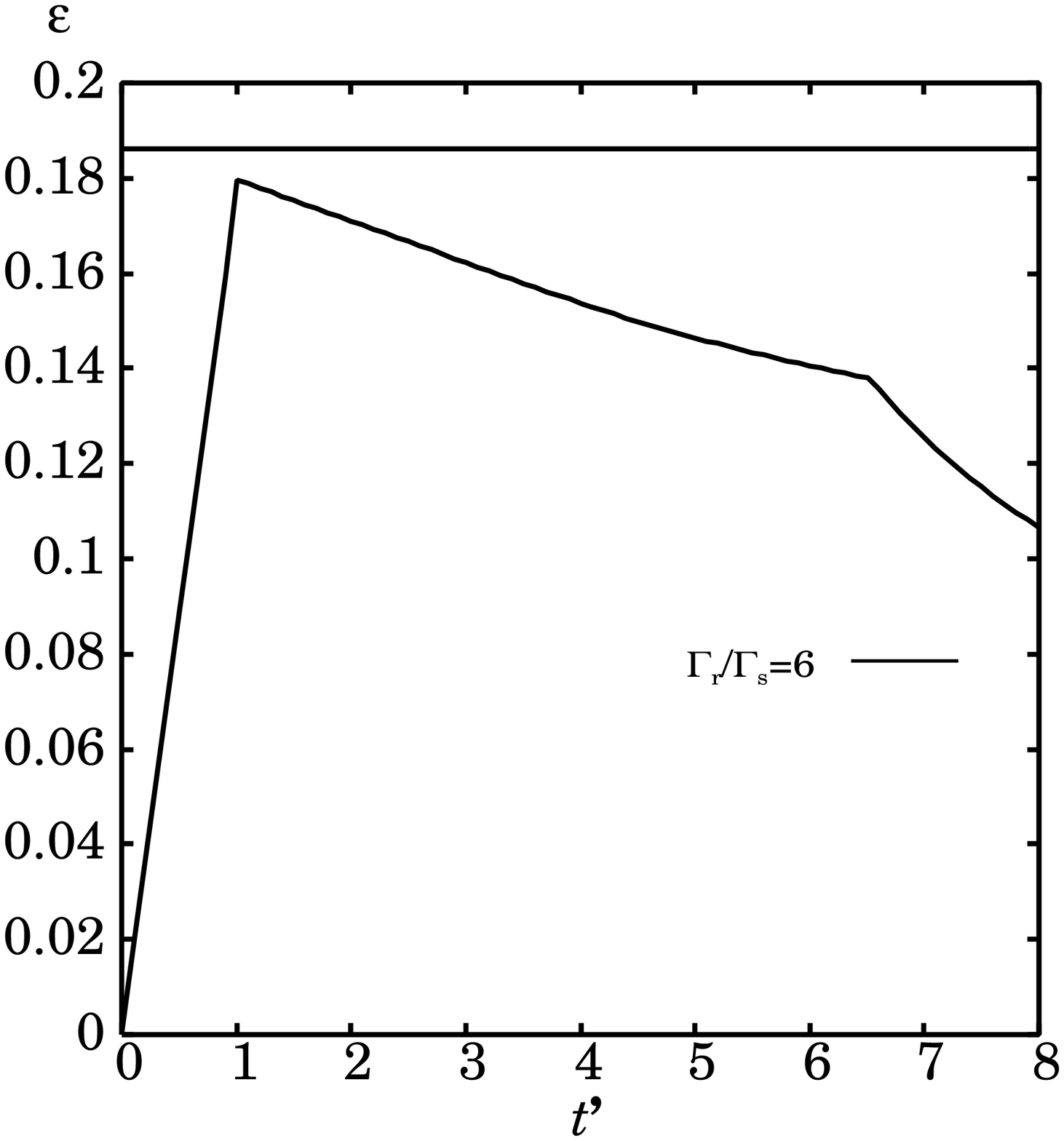}
\caption
%
{Time evolutions of the  
conversion efficiency defined by Eq. (\ref{eq:eff})
for the ``equal $\rho$'' case.
During shock propagations, 
$\epsilon$ 
approaches the two-point-mass estimate given by
Eq. (\ref{eq:eff-ana}) for each case.
}\label{fig:effrho}
\end{figure}
%


\end{document}